\documentclass[11pt,draftclsnofoot,peerreviewca,letterpaper,onecolumn]{IEEEtran}
\usepackage{cite}

\usepackage{graphicx,color,epsfig,rotating,subfigure}
\usepackage{amsfonts,amsmath,amssymb}
\usepackage{algorithm}
\usepackage{subfigure}
\usepackage{algpseudocode}

\long\def\comment#1{}

\newfont{\bbb}{msbm10 scaled 700}

\newfont{\bb}{msbm10 scaled 1100}
\newcommand{\CC}{\mbox{\bb C}}
\newcommand{\PP}{\mbox{\bb P}}
\newcommand{\RR}{\mbox{\bb R}}

\newcommand{\EE}{\mbox{\bb E}}

\newcommand{\hv}{{\bf h}}

\newcommand{\xv}{{\bf x}}
\newcommand{\yv}{{\bf y}}

\newcommand{\Am}{{\bf A}}

\newcommand{\Hm}{{\bf H}}
\newcommand{\Id}{{\bf I}}

\newcommand{\Ac}{{\cal A}}
\newcommand{\Bc}{{\cal B}}
\newcommand{\Cc}{{\cal C}}
\newcommand{\Dc}{{\cal D}}
\newcommand{\Ec}{{\cal E}}
\newcommand{\Fc}{{\cal F}}

\newcommand{\Sc}{{\cal S}}

\newcommand{\diag}{{\hbox{diag}}}
\renewcommand{\det}{{\hbox{det}}}
\newcommand{\trace}{{\hbox{tr}}}

\newcommand{\SNR}{{\sf SNR}}
\newcommand{\INR}{{\sf INR}}

\newcommand{\eqdef}{\stackrel{\Delta}{=}}

\newcommand{\herm}{{\sf H}}

\newtheorem{definition}{Definition}
\newtheorem{theorem}{Theorem}
\newtheorem{lemma}{Lemma}

\newtheorem{remark}{Remark}

\begin{document}

\title{On the Performance of Optimized Dense Device-to-Device Wireless Networks}
\author{\authorblockN{Song-Nam~Hong~\IEEEmembership{Student Member,~IEEE,}
        and~Giuseppe~Caire~\IEEEmembership{Fellow,~IEEE}}
\authorblockA{Department of Electrical Engineering, University of Southern California, Los Angeles, CA, USA}}

\maketitle

\begin{abstract}
We consider a device-to-device wireless network where $n$ users are densely deployed in a squared planar region and communicate with each other without the help of a wired infrastructure. For this network, we examine the three-phase hierarchical cooperation scheme originally proposed by
Ozgur, Leveque and Tse, and the two-phase improved hierarchical cooperation scheme successively proposed by Ozgur and Leveque based on the concept of {\em network multiple access}.  Exploiting recent results on the optimality of ``treating interference as noise'' in Gaussian interference channels,
we optimize the achievable average per-link rate and not just its scaling law (as a function of $n$).
In addition, we provide further improvements on both the previously proposed hierarchical cooperation schemes by a more efficient use of
TDMA and spatial reuse.

Thanks to our explicit achievable rate expressions, we are able to compare hierarchical cooperation scheme with multihop routing
(i.e., decode-and-forward relaying), where the latter can be regarded as the ``current practice'' of device-to-device infrastructureless wireless networks.
Our results show that the improved and optimized hierarchical cooperation schemes yield very significant rate gains over multihop routing in realistic conditions of
channel propagation exponents, signal to noise ratio, and number of users. This sheds light on the long-standing question
about the real advantage of hierarchical cooperation scheme over multihop routing beyond the well-known scaling laws analysis. In contrast, we also show that our rate optimization is non-trivial,
since when hierarchical cooperation is applied with off-the-shelf choice of the system parameters, no significant rate gain with respect to multihop routing is achieved.
We also show that for large pathloss exponent (e.g., $\alpha = 7$) the sum rate is a nearly linear function of the number of users $n$
in the range of networks of practical size ($n \leq 10^5$).  This also sheds light on a long-standing dispute on the effective achievability of
linear sum rate scaling with hierarchical cooperation. Finally, we notice that the achievable sum rate
for large $\alpha$ is much larger than for small $\alpha$ (e.g., $\alpha = 4$). This suggests that hierarchical cooperation scheme may be a very effective approach for networks operating at mm-waves, where the pathloss exponent is generally large.
\end{abstract}

\clearpage

\begin{IEEEkeywords}
Wireless network capacity, Dense device-to-device wireless networks, Hierarchical cooperation, Scaling laws.
\end{IEEEkeywords}

\section{Introduction}\label{sec:Intro}

Although it is extremely hard to characterize exactly the capacity of wireless networks, significant progress has been made towards the understanding of their theoretical limits. In \cite{Avestimehr}, the capacity of multiple multicast wireless network is approximately determined within a constant additive gap that depends on the number of nodes in the network but not one the signal-to-noise ratio (SNR) and on the channel coefficients. Also, for multiple flows over a single hop, capacity approximations were obtained in the form of degrees of freedom (DoF), generalized DoF (GDoF), and $O(1)$-gap bounds \cite{Cadambe08,Gou09,Jafar10}.  However, the multiflow-multihop case remains widely unsolved in general.
Scaling laws provide a useful way to characterize the behavior of the capacity of such networks when the number of nodes becomes large.
Initiated by Gupta and Kumar's seminal work \cite{Gupta}, extensive studies in the past decade have
made significant progress in the understanding of the scaling laws of such {\em large wireless networks}.
The well-known {\em multihop routing} (aka, decode and forward relaying) yields a sum capacity that scales as
$\Theta(\sqrt{n})$ \cite{Gupta}. In \cite{Ozgur}, Ozgur, Leveque and Tse proposed a cooperative architecture named {\em hierarchical cooperation}
that combines local communication and long-range cooperative MIMO communication.
Applying $t$ stages of the basic cooperative scheme to a dense network with $n$ users
in a hierarchical architecture, a capacity scaling of $\Theta(n^{\frac{t}{t+1}})$ was shown to be achievable.
Therefore, for any $\epsilon > 0$, a scaling $\Theta(n^{1-\epsilon})$ is achievable for sufficiently large $t$.
While the result of \cite{Gupta} holds for both a ``physical model'', that considers the actual standard communication-theoretic
complex baseband signal at each node receiver, and a ``protocol model'', that considers interference as a distance-based link conflict condition,
the result of  \cite{Ozgur} depends critically on modeling the channel coefficient between any two clusters of transmitting and receiving users as a full-rank
matrix, due to independent fading coefficients between different antenna pairs.  An interesting dichotomy appeared when in \cite{Franceschetti} Franceschetti, Migliore and Minero showed that the capacity is fundamentally
limited to scale as $\Theta(\sqrt{n})$. Instead of {\em assuming} an independent fading model, they
started from Maxwell's equations and counted the number of independent electromagnetic modes
that can flow across a cut separating two regions of the network, and combined this counting argument with a standard information theoretic cut-set bound.
This debate was resolved independently in \cite{Lee} and \cite{Ozgur1}, by showing that both results are correct and they are applicable
in different operating regimes of the network. Summarizing, they concluded that linear scaling is achievable if $n \leq \frac{\sqrt{\Ac}}{\lambda}$
where $\Ac$ denotes the area of network and $\lambda$ denotes the wavelength of the operating carrier frequency.
Consider for example a network of area $\Ac = 1$ km$^2$, operating at carrier frequency of 3 GHz ($\lambda = 0.1$m).
In this case, for $n \leq 10^4$ we would expect that the hierarchical cooperation architecture of \cite{Ozgur} yields a linear scaling of the sum capacity.
Such ``linear scaling'' of the network capacity with the number of users $n$ is the holy grail of large wireless networks
since it yields constant average rate per source-destination pair in the case where sources and destinations are randomly
selected such that their distance is $O(1)$. In turn, this implies that the network, in the linear scaling regime,
is ``scalable'' since the rate per end-to-end communication session does not vanish as the number of
users grows.

While scaling law analysis yields nice and clean results, it is hard to tell how a network really performs in terms of rates,
since such type of analysis typically fails to characterize the constants of the leading term in $n$ versus the next significant terms.
As a matter of fact, it has been often questioned whether hierarchical cooperation scheme can achieve significant rate gains over multihop routing for networks of
practical size and in realistic conditions of pathloss exponent and signal to noise ratio.

With this question in mind, the purpose of this paper is twofold.
First, we derive an achievable sum rate (not just a scaling law) for hierarchical cooperation scheme. In particular, we consider both the originally proposed hierarchical cooperation scheme in \cite{Ozgur} and the improved hierarchical cooperation scheme proposed successively by Ozgur and Leveque in \cite{Ozgur3}, based on the concept of {\em network multiple access}. We also provide improvements of these schemes through a better use of TDMA and spatial reuse.
Second, we optimize all such hierarchical cooperation schemes on the basis of the obtained sum rate expression, by appropriately choosing the transmit power,
the spatial reuse factor, and the quantization distortion level at the cooperative receivers.  Based on this optimization, we can provide
a concrete quantitative comparison between hierarchical cooperation scheme and multihop routing for networks of finite size. Our result confirm that
optimized hierarchical cooperation schemes can provide very significant rate gains over multihop routing in realistic conditions of
channel propagation exponents, signal to noise ratio, and number of users.
In contrast, we also show that careful system optimization of hierarchical cooperation is necessary and non-trivial,
since when hierarchical cooperation is applied with off-the-shelf choice of the system parameters, no significant rate gain with respect to multihop routing is achieved.
We also show that for large pathloss exponent (e.g., $\alpha = 7$) the sum rate is a nearly linear function of the number of users $n$
in the range of networks of practical size ($n \leq 10^5$), thus
shedding light on the long-standing question on the effective achievability of
linear sum rate scaling with hierarchical cooperation scheme in practice. Finally, we notice that the achievable sum rate
for large $\alpha$ is much larger than for small $\alpha$ (e.g., $\alpha = 4$). This suggests that hierarchical cooperation scheme is
especially suited for networks operating at millimeter-waves, where the pathloss exponent is generally large.

Going back to scaling laws, we confirm and improve the previous result in \cite{Ghaderi} where, based on a more refined scaling law analysis,
it was found that the optimal number of hierarchical stage is $O(\sqrt{\log{n}})$.
Since our analysis is based on achievable rate expressions and not only on scaling laws, we are able to provide a more
precise quantitative characterization of the optimal number of stages for given finite $n$.
For example, we show that for $n \leq 10^4$ (as in the above example)
the optimal number of stages $t$ is less than $3$ (see Fig.~\ref{opt_t}).\footnote{Notice that $n = 10^4$ nodes in 1 km$^2$ is already a very high density of nodes in practical sensor networks and tactical device-to-device networks, and even increasing the carrier frequency from 3 GHz to 30 GHz, i.e., pushing the upper bound on the network size to $n \leq 10^5$, yields an optimal number of stages $t$ not larger than 4.
Hence, our conclusion stands even considering networks that operate in the mm-wave range  \cite{Daniels, rappa, rappa-mag}.}

Notice that the linear scaling in \cite{Ozgur} is obtained by {\em first} letting $n \rightarrow \infty$ to get the scaling
law of a single stage of the hierarchical cooperation scheme, and {\em then} letting $t \rightarrow \infty$ to achieve the linear scaling of the sum rate.
Notice also that the order of the limits here is critical. In fact, letting
$n \rightarrow \infty$ is not possible because it violates the
``electromagnetic propagation'' dimensionality bound $n \leq \frac{\sqrt{\Ac}}{\lambda}$ said above.
In contrast, the behavior of a sequence of networks of increasing size $n$ should be characterized by letting $n \rightarrow \infty$ and
optimizing $t$ a as a function of $n$, for each finite $n$. In light of our results on the optimal number of hierarchical cooperation stages $t$ (see also  \cite{Ghaderi}),
one is tempted to conclude that linear scaling of the sum rate is not feasible by hierarchical cooperation in any regime.
In contrast, thanks to our explicit rate analysis, we can conclude that suitably optimized hierarchical cooperation schemes {\em effectively} achieve
a near-linear scaling of the sum rate in the range of practical network sizes, when the pathloss exponent is not too small.

{\bf System model:}
We consider a network deployed over a unit-area squared region and formed by $n$ nodes placed
on a regular grid with minimum distance $1/\sqrt{n}$. The grid topology captures the essence of the problem while avoiding some
technicalities due to the random placement of nodes.
The network consists of $n$ source-destination pairs, such that
each node is both a source and a destination, and pairs are selected at random over the set of $n$-permutation $\pi$
that do not fix any element (i.e., for which $\pi(i) \neq i$ for all $i = 1,\ldots, n$). We focus on max-min fairness, such that all
source-destination pairs wish to communicate at the same rate.
The channel coefficient between a transmitter node $k$ and a receive node $\ell$ at distance $r_{\ell k}$ is given by
$h_{\ell k} = r_{\ell k}^{-\alpha/2}\exp{(j\theta_{\ell k})}$, where $\alpha$ denotes the path-loss exponent
and $\theta_{\ell k}\sim\mbox{Unif}(0,2\pi]$ denotes a random i.i.d. phase.
This independent ``phase fading'' model is the same assumed in \cite{Ozgur,Ozgur3,Ghaderi}.

{\bf Paper organization:}
In Section~\ref{sec:CTS}, we derive an achievable sum rate for the basic three-phase cooperative transmission scheme of \cite{Ozgur}.
We also optimized such rate by appropriately choosing the transmit power, reuse factor, and quantization distortion level of the scheme.
This result is extended to the hierarchical cooperation architecture in Section~\ref{sec:H-CTS}.
 We also consider various improvements on the basic hierarchical cooperation scheme of \cite{Ozgur}, where we employ more efficiently the TDMA during the local communication phases and consider the local communication phase as a network multiple access problem as in \cite{Ozgur3}.
In Section~\ref{sec:comp}, we compare hierarchical cooperation scheme with multihop routing and show that the optimized hierarchical cooperation scheme largely outperforms multihop routing, even for a
moderately large network size. Some concluding remarks are provided in Section~\ref{sec:conclusion}.

\begin{figure}
\centerline{\includegraphics[width=5cm]{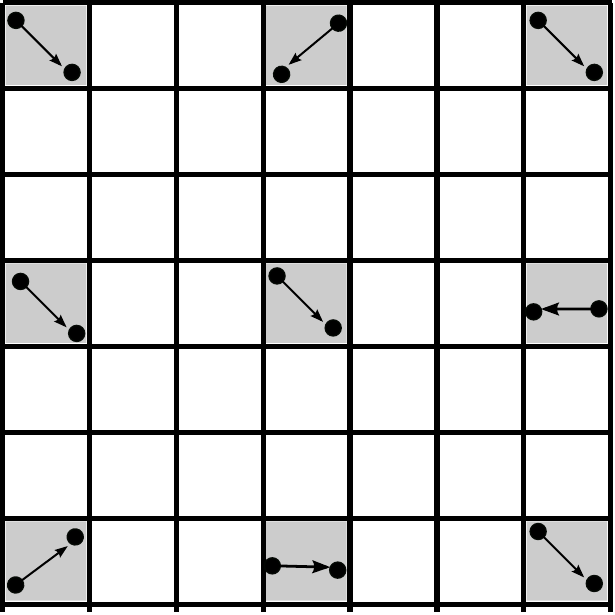}}
\caption{Grouping of interfering clusters in the TDMA scheme with reuse factor $L=3$.
Each square represents a cluster and the gray squares represent the concurrent transmitting clusters.}
\label{TDMA}
\end{figure}

\section{Cooperative Transmission Scheme}\label{sec:CTS}

\begin{figure}
\centerline{\includegraphics[width=14cm]{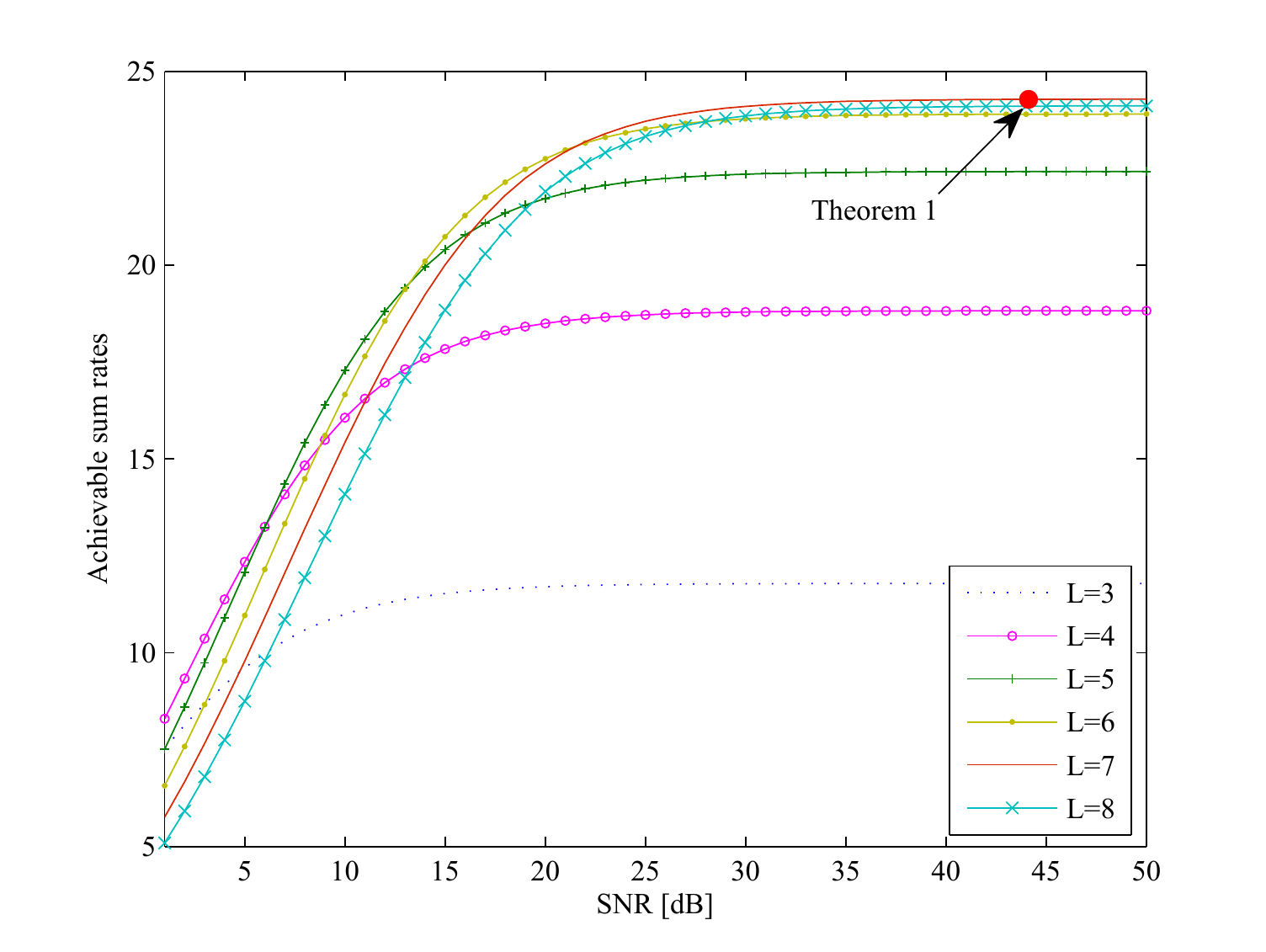}}
\caption{Achievable sum rates of the cooperative transmission scheme as a function of reuse factor $L$ when pathloss exponent $\alpha=3$ and network size $n=10^{4}$. }
\label{opt_TDMA}
\end{figure}

\begin{figure}
\centerline{\includegraphics[width=14cm]{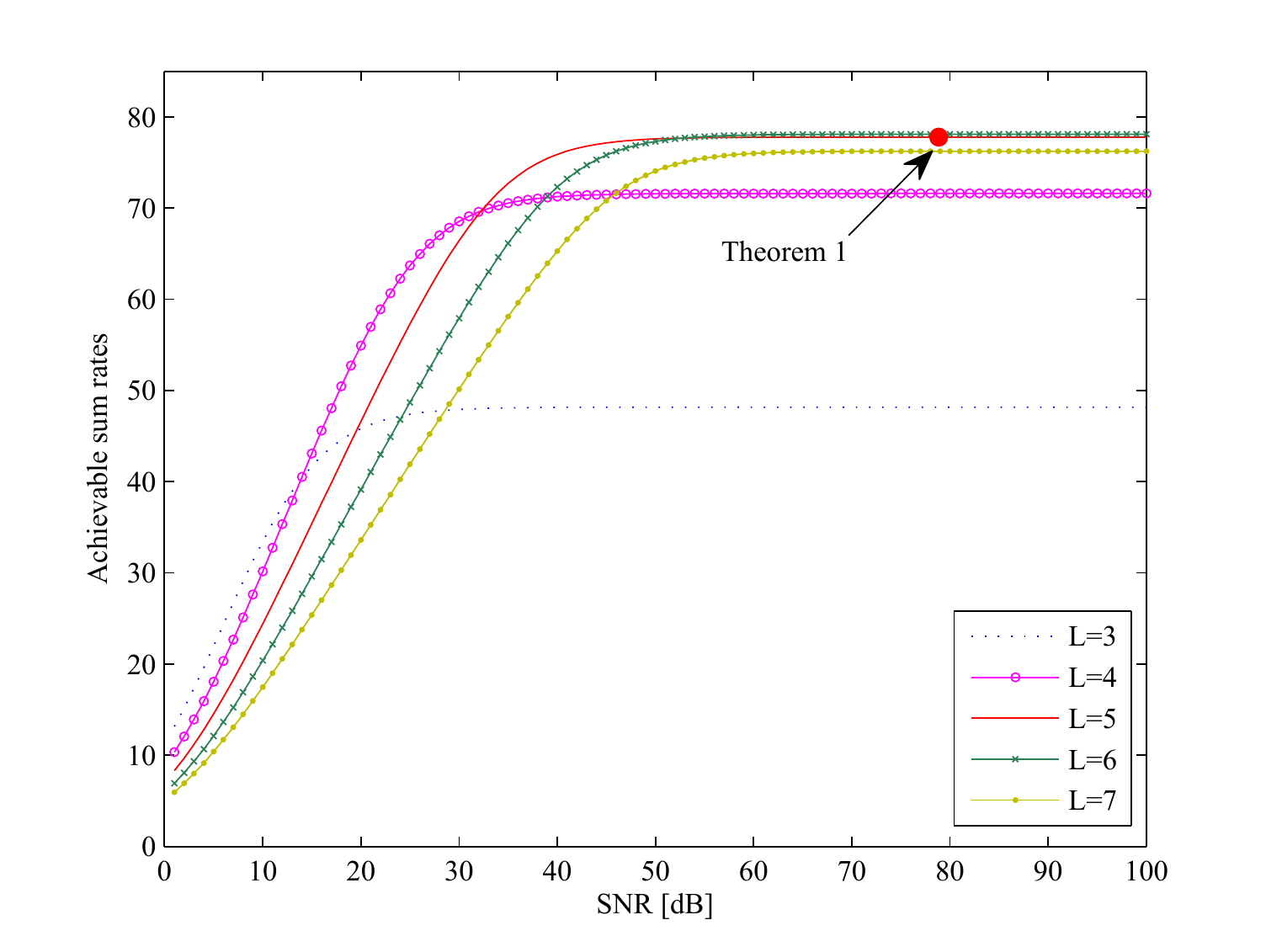}}
\caption{Achievable sum rates of the cooperative transmission scheme as a function of reuse factor $L$ when pathloss exponent $\alpha=7$ and network size $n=10^{4}$.}
\label{opt_TDMA1}
\end{figure}

In this section, we optimize the cooperative transmission scheme proposed in \cite{Ozgur}
with respect to the achievable sum rate.
We let $R_{{\rm c}}(\alpha)$ denote the common message {\em coding rate} for all users, expressed in bits per codeword symbol.
The scheme delivers $n$ messages using a certain number of time slots, each of which corresponds to the duration of a codeword.
Hence, the network sum {\em throughput} is given by \[R_{\rm sum}(n,\alpha) = R_{{\rm c}}(\alpha)\mbox{T}(n,\alpha),\] where
\[  \mbox{T}(n,\alpha)= n/\mbox{(required number of time slots)}, \]
is the number of delivered source-destination messages per time slot ratio (referred to as {\em packet throughput}
in the following).

The network is divided into $n/M$ clusters of $M$ nodes each. The cooperative transmission scheme consists of three phases:
\begin{itemize}
\item {\bf Phase 1: Information dissemination.}  Each source distributes $M$ distinct sub-packets of its message to the $M$ neighboring nodes in the same cluster. One transmission is active per each cluster, in a round robin fashion, and clusters are active simultaneously
in order to achieve  some spatial spectrum reuse. The inter-cluster interference is controlled by the reuse factor $L$, i.e., each cluster has one transmission opportunity every $L^2$ time slots (see Fig.~\ref{TDMA}).

\item {\bf Phase 2: Long-range MIMO transmission.}  One cluster at a time is active, and when a cluster is active it operates as a single $M$-antenna MIMO transmitter, sending $M$ independently encoded
data streams to a destination cluster. Each node in the cooperative receiving cluster stores its own received signal.
\item {\bf Phase 3: Cooperative reception.} All receivers in each cluster share their own received and quantized signals so that each destination in the cluster decodes its intended message on the basis of the (quantized) $M$-dimensional observation. Quantization and binning (or random hashing
of the quantization bits onto channel codewords) is used in this phase, which is a special case of the general quantize-remap-and-forward (QMF) scheme (also referred to as noisy network coding)
for wireless relay networks \cite{Avestimehr, Lim}. Each destination performs joint typical decoding to obtain its own desired message
based on the quantized signals (or bin indices).
\end{itemize}

The parameters that must be optimized in the cooperative transmission scheme are the cluster size $M$,
the node transmit power $\SNR$, and  reuse factor $L$. Regarding the transmit power, it is assumed that $\SNR$ can be chosen arbitrarily with a uniform
bound $\SNR \leq \SNR_{\max}$ where the latter is a fixed arbitrarily large constant that does not scale with $n$. As the result of such optimization, we have:

\begin{theorem}\label{thm:sum-rate} For network size $n$ and path-loss exponent $\alpha$, the cooperative transmission scheme achieves the sum rate of
\begin{equation*}
R_{\rm sum}(n,\alpha) = \log\left(1+\frac{\SNR}{1+P_{I}}\right)\frac{\sqrt{n}}{2\sqrt{2}L(\SNR)} \label{eq:sum1}
\end{equation*}where
\begin{eqnarray}
\SNR &=& 2^{2(3+\alpha/\ln{2})}\\
L(\SNR) &=& \left\lceil\sqrt{\SNR}^{1/\alpha}+1\right\rceil\\
P_{I} &=&\sum_{i=1}^{\sqrt{n}} 8i\SNR (L(\SNR)i-1)^{-\alpha}.\label{eq:IntPower}
\end{eqnarray}
\end{theorem}
\begin{IEEEproof} See Sections~\ref{subsec:LC}, \ref{subsec:MIMO}, and~\ref{subsec:ST1}.
\end{IEEEproof}


Theorem \ref{thm:sum-rate} implies that all sources can reliably transmit their messages at rate
$R_{{\rm c}}(\alpha) \approx\log{\sqrt{\SNR}}$ over the $2\sqrt{2}L(\SNR)\sqrt{n}$ time slots. Notice that despite the fact we let $\SNR_{\max}$ to be an arbitrarily large constant, the optimal SNR depends only on the pathloss $\alpha$ and it is generally not too large. This is because there is tension between the transmit power of each local link and the reuse factor necessary to keep inter-cluster interference under control.
The optimal transmit power is determined in Section~\ref{subsec:ST1} (Theorem~\ref{thm:sum-rate}), as a result of this tradeoff.
For comparison, notice that in the original scheme of \cite{Ozgur} the reuse factor is fixed to $L=3$.
Figs.~\ref{opt_TDMA} and~\ref{opt_TDMA1} show that our optimized scheme provides a substantial gain over the scheme of \cite{Ozgur}.

\begin{remark} By considering only the dominant interfering terms at each receiver,
the total interference power is well-approximated by:
\begin{equation}
P_{I} = 8\SNR(L(\SNR)-1)^{-\alpha}.
\end{equation} With the approximation $L(\SNR) \approx \sqrt{\SNR}^{1/\alpha} + 1$, the achievable sum rate in Theorem~\ref{thm:sum-rate} can be
simplified as
\begin{equation}
R_{\rm sum}(n,\alpha) \approx \frac{\alpha \sqrt{n}}{(2\sqrt{2}\ln{2})2^{3/\alpha + 1/\ln{2}}}.\label{eq:app-sumrate}
\end{equation}
From (\ref{eq:app-sumrate}), we observe that the sum rate grows almost linearly with the pathloss exponent $\alpha$.
Thus, the optimized cooperative transmission scheme can provide a higher rate when the system operates
at high frequencies (e.g., mm-waves) due to the fact that at those frequencies the pathloss exponent
becomes large \cite{rappa-mag, Yanikomeroglu}.
 \hfill $\lozenge$
\end{remark}

\subsection{Information dissemination by TDMA and spatial reuse}\label{subsec:LC}

Phases 1 and 3 of the scheme employ intra-cluster TDMA and spatial reuse across the clusters. Therefore, there is inter-cluster interference between simultaneously active clusters. Since only one source-destination pair is active per cluster and there are $n/M$ cluster, we are in the presence of a $n/M$ user interference channel. For such channel, advanced coding schemes such as message splitting and successive interference cancellation \cite{Han,Etkin}, interference alignment \cite{Cadambe}, and structured coding \cite{Ordentlich}, require extensive channel state information at both transmitters and receivers.
On the other hand, it has been recently recognized in \cite{Geng} that there exists a regime of the interference channel gains for which ``Gaussian" single-user capacity achieving codes and {\em treating interference as noise} (TIN) is information-theoretically optimal (within a constant gap).
TIN is most attractive in practice since it requires standard  codes and minimum
distance decoders. Hence, we design our optimized scheme such that the network operates in the TIN-optimal regime. For convenience, we recall here the main result of \cite{Geng}:
\begin{theorem}[\cite{Geng}]\label{thm:TIN} For $n$-user interference channel, if the following condition is satisfied, then treating interference as noise (TIN) can achieve the whole capacity region to within a constant gap of $\log(3n)$:
\begin{equation}
\SNR_{i} \geq \max_{j\neq i}\INR_{ij} \cdot \max_{k \neq i}\INR_{ki}, \forall i = 1,\ldots, n,
\end{equation}  where $\SNR_{i}  = \frac{P|h_{ii}|^2}{N_{0}}$ and $\INR_{ij} = \frac{P|h_{ij}|^2}{N_{0}}$ denote the signal-to-noise ratio of user $i$ and the interference-to-noise ratio of source $j$ at destination $i$, respectively. \flushright$\blacksquare$
\end{theorem}

In order to ensure that the interference channel induced by the local communication phases operates in the regime for which TIN is (near) optimal,
we can choose appropriately the reuse factor $L$.
We first determine the transmit power $P$ according to the cluster area $\Ac$ and the pathloss exponent $\alpha$
as $P=\SNR\Ac^{\alpha/2}$. This choice yields $\SNR_{i} \geq \SNR$ for all $i$.  Also, let $\INR$ denote the strongest interference power, i.e., $\INR = \max_{j\neq i}\INR_{ij} = \max_{k \neq i}\INR_{ki}$ where the last equality is due to the symmetric structure of network.
Considering the TDMA structure, we obtain that $\INR \leq (L-1)^{-\alpha}\SNR$. Due to symmetry, the optimality condition of TIN in Theorem~\ref{thm:TIN}
is satisfied if $\INR \leq \sqrt{\SNR}$. We can find $L$ to meet the above condition as \[L(\SNR) =  \left\lceil\sqrt{\SNR}^{1/\alpha}+1\right\rceil.\]
Then, the local communication rate of  $R^{(1)}(\SNR) = \log\left(1+\frac{\SNR}{1+P_{I}}\right)$ is achievable by TIN, where
the inter-cluster total interference is upper bounded by $P_{I}$ in (\ref{eq:IntPower}).
Reliable local communication is ensured by letting:
\begin{equation}
R_{{\rm c}}(\alpha) \leq R^{(1)}(\SNR). \label{eq:rate-const1}
\end{equation} From Figs.~\ref{opt_TDMA} and~\ref{opt_TDMA1}, we observe that the achievable sum rate vs. $\SNR$ curves, for fixed values of $L$, are non-decreasing and saturate to an horizontal asymptote. We notice also that the pair of $L$ and $\SNR$ chosen according to the above TIN criterion yield a sum rate essentially equal to the maximum achievable rate, when maximizing also with respect to $L$, although, for fixed gap $\epsilon > 0$ from the maximum, a lower transmit power per node is generally possible. Since in our system optimization we are not concerned  in minimizing transmit power, we conclude that the proposed TIN criterion yields a useful and effective way to optimize the sum rate in closed-form, without the need of performing system simulation.

\subsection{Long-range MIMO communication and cooperative reception}\label{subsec:MIMO}

\begin{figure}
\centerline{\includegraphics[width=8cm]{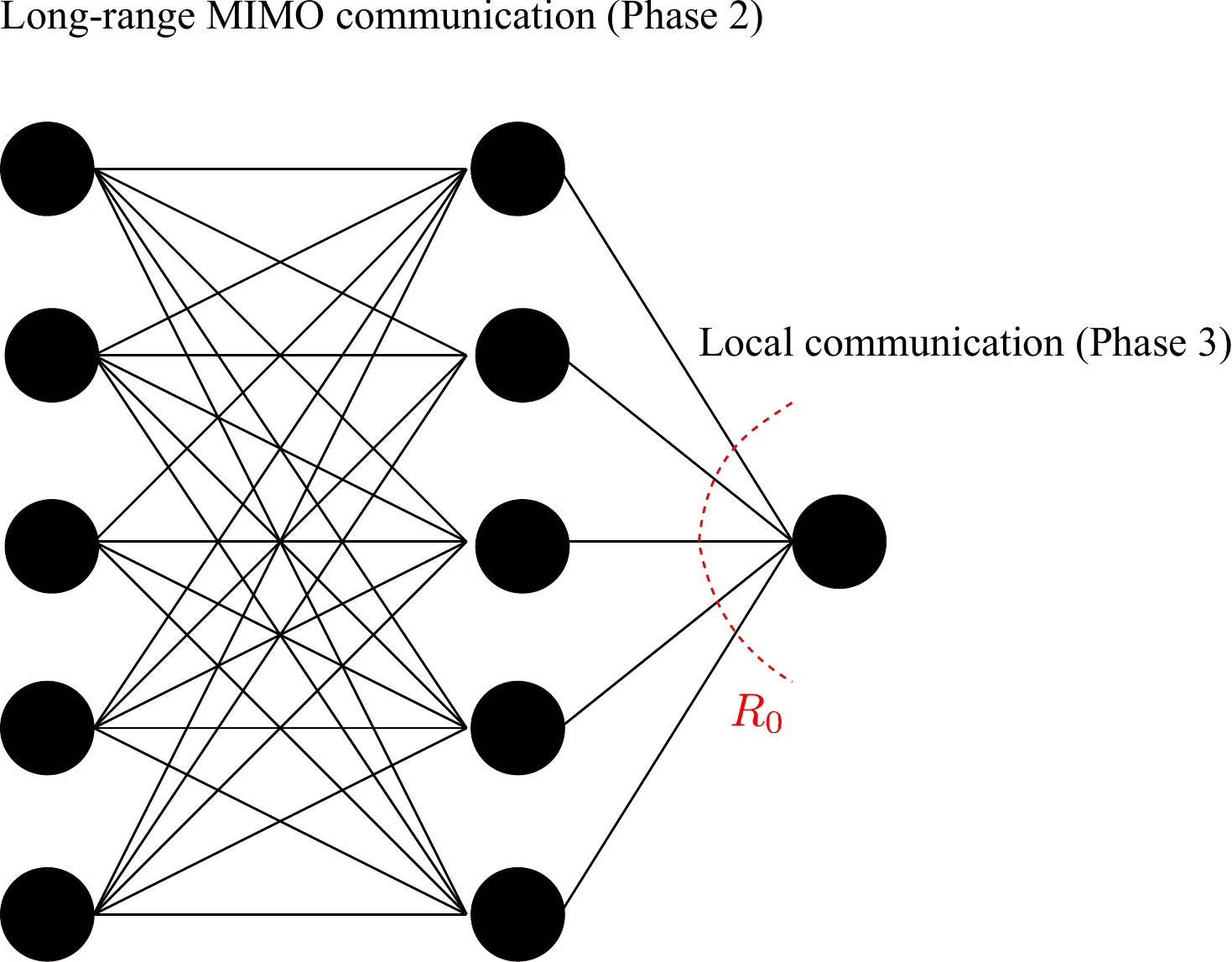}}
\caption{Distributed MIMO channel with finite backhaul capacity of rate $R_{0}$ where $R_{0}$ is determined by local communication rate.}
\label{DAS}
\end{figure}

By concatenating phases 2 and 3 of the cooperative scheme, we obtain an equivalent  distributed MIMO channel with finite backhaul capacity
of rate $R_{0}$  as shown in  Fig.~\ref{DAS},
where the $M$ transmit (resp., $M$ receiver) antennas correspond to the $M$ nodes in the source cluster (resp., destination cluster).
This model has been extensively studied in \cite{Sanderovich,Hong}. In particular, in \cite{Hong} we showed that
the capacity of this channel is almost achieved by either QMF or compute-and-forward (CoF) \cite{Nazer}, depending on
the channel coefficients and on the value of $R_0$.
Specifically, CoF can outperform QMF if the number of strong interferers at each receiver is relatively small with respect to the total number of nodes
in the network (i.e., {\em sparse network}) and the backhaul capacity is small (i.e., $R_{0} \leq 5$).
In this regime, the impact of the non-integer penalty of CoF is not severe, while QMF suffers from the quantization noise due to the small backhaul
capacity (see \cite{Hong} for details). In other regimes, QMF shows better performance than CoF because it can fully exploit the MIMO beamforming gain.
Since our model can be considered as a {\em dense network} (i.e., each receiver suffers from $n-1$ non-negligible interfering nodes),
here we choose to employ QMF for the long-range MIMO transmission. In QMF \cite{Hong}, receivers perform vector quantization of their received signal at some rate $R'\geq R_{0}$ where $R_{0}$ denotes a backhaul capacity (see Fig.~\ref{DAS}). They map the blocks of $nR'$ quantization bits into binary words of length $nR_{0}$ by using some randomized hashing function (notice that this corresponds to binning if $R' > R_{0}$), and let the destination perform joint decoding of all source messages based on the observations of all the (hashed) quantization bits. We remark that in \cite{Ozgur},  quantize-and-forward (QF), a simplified version of QMF without using binning (see \cite{Hong} for details), was used.

For the MIMO transmission (i.e., phase 2), the transmit power is given by
\[P_{\mbox{\tiny MIMO}}=\frac{\SNR'}{M}\Ac^{\alpha/2},\]
where $\SNR'$ can be arbitrary chosen with $\SNR'\leq \SNR_{\max}$ as in the local communication.
Using distance-dependent power control in order to eliminate the effect of the pathloss between the transmit and receiving clusters,
the channel matrix of the distributed MIMO channel is $\Hm \in \CC^{M \times M}$, with $(k,\ell)$-element given by $\exp(j\theta_{k\ell})$,
with $\theta_{k\ell} \sim \mbox{Unif}(0,2\pi]$. Let $N_{0}$ denote the variance of the additive noise plus inter-cluster interference.\footnote{Inter-cluster interference is zero in a single layer of hierarchical cooperation, but is non-zero when multiple stages are considered in hierarchical cooperation architecture.
Therefore, it is denoted here by $N_0$ not necessarily equal to the normalized noise level $1$, for the sake of generality.}
As in \cite{Ozgur}, the local communication of phase 3 can be expanded over $Q$ time slots for some integer $Q$, in order to obtain more flexibility in the
quantization rate of the underlying QMF scheme.  This yields the backhaul capacity of the ``equivalent'' model
as $R_{0} = QR^{(1)}(\SNR)$.  An optimal $Q$ will be chosen later on.

The computation of the rate achievable by QMF for the distributed MIMO channel with finite backhaul capacity is
generally difficult  since it involves a complicated combinatorial optimization  \cite{Sanderovich}.
So far, a closed-form expression was only available for the symmetric Wyner model \cite{Sanderovich}.
In this paper, we derive a closed-form expression of  the achievable rate for our model, exploiting the fact that, for large $n$,
the problem becomes symmetric although the matrix $\Hm$ is ``full'' and not tri-diagonal as in the Wyner model. Our result is based on
asymptotic {\em random matrix theory} and the submodular structure of the rate expression:

\begin{theorem}\label{thm:DASrate} For a distributed MIMO channel with backhaul capacity equal to $R_{0}$ and random i.i.d. channel
coefficients with zero mean and unit variance, QMF achieves the symmetric rate of
\begin{eqnarray}
R_{{\rm QMF}}(R_{0},N_{0},\SNR)
&=& \min\left\{R_{0}-\log\left(1+\frac{N_{0}}{\sigma_{q}^2}\right), \Cc\left(\frac{\SNR}{N_{0}+\sigma_{q}^2}\right)\right\}, \label{eq:QMF}
\end{eqnarray} for some quantization level $\sigma_{q}^2 \geq 0$, where
\begin{equation} \label{Cx}
\Cc(x)\eqdef 2\log\left(\frac{1+\sqrt{1+4x}}{2}\right)-\frac{\log{e}}{4x}(\sqrt{1+4x}-1)^2.
\end{equation}
\end{theorem}
\begin{IEEEproof}
See Appendix~\ref{sec:DAS-RATE}.
\end{IEEEproof}

Since the achievable rate in (\ref{eq:QMF}) is the minimum of two terms, where the first is an increasing function of $\sigma_{q}^2$ and the second
is a decreasing function of $\sigma_{q}^2$, the optimal value of $\sigma_{q}^2$ is attained by solving
\[R_{0}-\log\left(1+\frac{N_{0}}{\sigma_{q}^2}\right)= \Cc\left(\frac{\SNR}{N_{0}+\sigma_{q}^2}\right).\]
Letting
\[f(\sigma_{q}^2) \eqdef R_{0}-\log\left(1+\frac{N_{0}}{\sigma_{q}^2}\right) - \Cc\left(\frac{\SNR}{N_{0}+\sigma_{q}^2}\right),\]
we can find
\[\sigma_{q,{\rm min}}^2 = \frac{N_{0}}{2^{R_{0}} -1} \mbox{ and } \sigma_{q,{\rm max}}^2=\frac{N_{0}+\SNR}{2^{R_{0}}-1},\]
 such that $f(\sigma_{q,{\rm min}}^2) \leq 0$  and  $f(\sigma_{q,{\rm max}}^2) \geq 0$.
This is because $\sigma^2_{q,{\rm min}}$ makes the first term of the minimum in (\ref{eq:QMF}) zero and $\sigma^2_{q,{\rm max}}$
is the quantization level of QF,
which makes the second term to attain the minimum in (\ref{eq:QMF}). Using bisection method, we can quickly find an optimal quantization
level $\sigma^2_{q,{\rm opt}}$. This will be used in this paper to plot the achievable rates of QMF.

Putting together the MIMO rate constraint of Theorem \ref{thm:DASrate} with the rate achievable
in phase 1 (\ref{eq:rate-const1}), we find
\begin{equation*}
R_{{\rm c}}(\alpha) = \min\{R^{(1)}(\SNR), R_{{\rm QMF}}(QR^{(1)}(\SNR),1,\SNR'))\}.
\end{equation*}
Since there is no inter-cluster interference (i.e., $N_{0}=1$) in the MIMO communication phase 2,
we can find some finite value $\SNR'$ with $Q=1$ such that \[R^{(1)}(\SNR) \leq R_{{\rm QMF}}(R^{(1)}(\SNR),1,\SNR').\]
Then, we have that $R_{{\rm c}}(\alpha) = R^{(1)}(\SNR)$, where the optimal $\SNR$ will be determined in the next section.
In fact, we do not have to compute an exact explicit expression for the achievable rate of QMF
in this section but the QMF rate will be used in Section~\ref{sec:H-CTS} for the hierarchical cooperation architecture,
when we shall consider multiple stages of the three-phase cooperative scheme.

\begin{remark}
We observe that, by choosing $\SNR'=\SNR$, the achievable coding rate is about $R_{{\rm c}} = R^{(1)}(\SNR) - 0.5$ when using the optimal quantization level. This has about 0.5 bits improvement compared with noise-power
level quantization in \cite{Avestimehr}. The improvement becomes  larger when considering a hierarchical cooperation scheme with $t>1$ stages. In this case, QMF is applied to a multihop distributed MIMO channel consisting of $t$ relay stages (see Section~\ref{subsec:M-RATE}). Due to the accumulation of quantization noises, the rate of QMF decreases as $t$ grows. Remarkably, we observe that with noise-level quantization, the gap grows {\em linearly} with the number of stages $t$ while with optimal quantization, the gap grows {\em logarithmically} with the number of stages $t$. This shows that indeed optimizing the quantization level can significantly improve the rate of QMF. Analogous results have been shown in \cite{Chern,Kolte,Hong-full,Hong-ITW2015} for other Gaussian networks.
\hfill $\lozenge$
\end{remark}

\subsection{Achievable sum rate}\label{subsec:ST1}

In order to derive an achievable sum rate, we will compute the packet throughput $\mbox{T}(n,\alpha)$.
As anticipated before, in the cooperative scheme each source transmits $M$ distinct sub-packets of its message to the intended destination.
To transmit overall $nM$ sub-packets (in the whole network), phase 1 requires
$(L(\SNR)M)^2$ time slots, phase 2 requires $n$ time slots, and phase 3 requires $Q(L(\SNR)M)^2$ time slots.
Hence, we have
\[\mbox{T}(n,\alpha) = \frac{Mn}{(Q+1)(L(\SNR)M)^2+n}.\]
Since the coding rate $R^{(1)}(\SNR)$ is independent of $M$, we can find the optimal cluster size $M$
by treating $M$ as a continuous variable and solving $\frac{d \mbox{T}(n,\alpha)}{ d M} = 0$. This yields
\[M = \frac{\sqrt{n}}{L(\SNR)\sqrt{1+Q}}.\]
Then, the packet throughput is obtained as
\[\mbox{T}(n,\alpha) = \frac{\sqrt{n}}{2L(\SNR)\sqrt{1+Q}},\] and accordingly, the achievable sum rate is given by
\[R_{{\rm sum}} (n,\alpha) = R^{(1)}(\SNR)\frac{\sqrt{n}}{2L(\SNR)\sqrt{1+Q}}.\]
Next, we will optimize the transmit power $\SNR$ to maximize the above sum rate.
To make the problem tractable, we use the approximations $L(\SNR) \approx \sqrt{\SNR}^{1/\alpha}$ and $R^{(1)}(\SNR) \approx \log(\sqrt{\SNR}/8)$.
Then, the sum rate is approximated by
\[\tilde{R}_{{\rm sum}} (n,\alpha) =\log\left(\frac{\sqrt{\SNR}}{8}\right)\frac{\sqrt{n}}{2\sqrt{2}\sqrt{\SNR}^{1/\alpha}},\]
where  $Q=1$ is chosen because of the argument given in Section \ref{subsec:MIMO}.
Differentiating and solving $\frac{d \tilde{R}_{{\rm sum}} (n,\alpha)}{d \SNR}=0$, we find that the optimal transmit power is given by
\[\SNR= 2^{2(3+\alpha/\ln2)}.\]
This concludes the proof of Theorem~\ref{thm:sum-rate}.


\section{Optimizing the hierarchical cooperation architecture}\label{sec:H-CTS}

The basic hierarchical cooperation architecture was proposed in \cite{Ozgur} by employing the three-phase
cooperative transmission scheme reviewed and analyzed in Section~\ref{sec:CTS} as a recursive building block,
applied for local communication of a higher stage, i.e., at a larger space scale
in the network. In the overall hierarchical cooperation scheme, we use the symmetric coding rate $R_{{\rm c}}(\alpha)$ regardless of the number of
hierarchical stages $t$. Based on Section~\ref{sec:CTS}, we choose
\begin{eqnarray*}
P_{i} &=& \SNR \Ac_{i}^{\alpha/2}\\
L &=& \left\lceil \sqrt{\SNR}^{1/\alpha} +1 \right \rceil\\
\SNR&=&2^{2(3+\alpha/\ln{2})},
\end{eqnarray*} for stages $i=1,\ldots,t$, where $\Ac_{i}$ denotes the cluster area of stage $i$.
Notice that these choices guarantee that, regardless of the hierarchical stage $i$, the received power of inter-cluster interference is
upper bounded by $P_{I}$ given in (\ref{eq:IntPower}). For the MIMO communication phase, we choose
transmit power \[P_{\mbox{{\tiny MIMO}},i}=\frac{\SNR}{M} \Ac_{i}^{\alpha/2},\] which also makes the interference power to be not larger than
$P_{I}$.
 As a matter of fact, the original hierarchical cooperation scheme proposed in \cite{Ozgur} can be improved. One improvement (here referred to as Method 3)
was proposed in \cite{Ozgur3}, where the local communication phase is formulated as a network multiple access problem
instead of decomposing it into a number of unicast network problems (see Section~\ref{subsubsection:approach3}).
In this paper, we further improve the throughput of both Method 1 and Method 3 by using a more efficient TDMA scheduling.
The resulting enhanced schemes are referred to as Method 2 and Method 4, respectively. The main result of this section is:

\begin{theorem}\label{thm:H-sumrate}
For network size $n$ and path-loss exponent $\alpha$, hierarchical cooperation scheme (in the various forms summarized above)
with $t\geq 2$ stages achieves the sum rate of
\begin{equation*}
R_{{\rm sum}}^{(t)}(n,\alpha)=\left\{
                                \begin{array}{ll}
                                  R_{{\rm c}}(\alpha)\frac{n^{\frac{t}{t+1}}}{(1+t)(L\sqrt{3})^{t}}, & \hbox{Method 1} \\
                                  R_{{\rm c}}(\alpha)\frac{n^{\frac{t}{t+1}}}{(1+t)L^{\frac{2t}{t+1}}\sqrt{3}^{t}}, & \hbox{Method 2} \\
                                  R_{{\rm c}}(\alpha)\frac{n^{\frac{t}{t+1}}}{(1+t)L^{t}(3\cdot2^{t-1})^{\frac{t}{2(t+1)}}}, & \hbox{Method 3} \\
                                  R_{{\rm c}}(\alpha)\frac{n^{\frac{t}{t+1}}}{(1+t)L^{\frac{2t}{t+1}}(3\cdot2^{t-1})^{\frac{t}{2(t+1)}}}, & \hbox{Method 4,}
                                \end{array}
                              \right.
\end{equation*} where $L= \left\lceil 2^{(3+\alpha/\ln{2})/\alpha} +1 \right\rceil$, and $R_{{\tiny c}}(\alpha)$ is determined in Section~\ref{subsec:M-RATE} and is given in Fig.~\ref{MRates} for $3\leq \alpha\leq 11$.
\end{theorem}
\begin{IEEEproof}
See Sections~\ref{subsec:M-RATE} and~\ref{subsec:ST}.
\end{IEEEproof}

\begin{figure}
\centerline{\includegraphics[width=14cm]{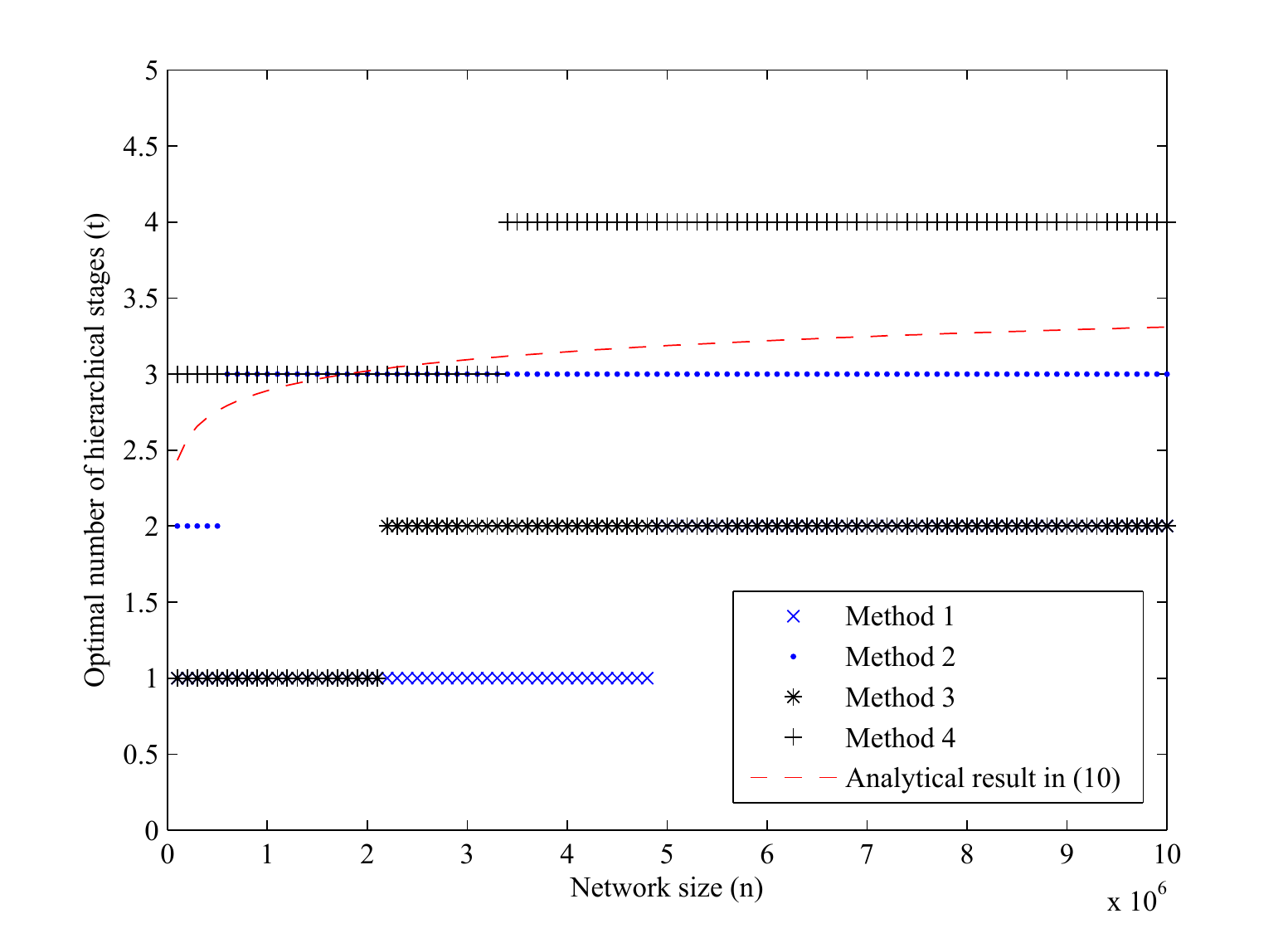}}
\caption{Optimal number of hierarchical stages $t$ as a function of network size $n$, when pathloss exponent $\alpha = 7$.}
\label{opt_t}
\end{figure}
Notice that even for $t=1$ (single stage), the sum rate in Theorem~\ref{thm:H-sumrate} does not reduce to the previous result of Theorem~\ref{thm:sum-rate} since in this case, a higher coding rate can be chosen because there is no inter-cluster interference in the MIMO communication phase. Instead, the coding rate in Theorem~\ref{thm:H-sumrate} is chosen in order to meet the most stringent constraint for reliable communication with an arbitrary number of stages $t$, i.e., including the case where the MIMO communication phase of stages $i<t$ suffers from inter-cluster interference.

From Theorem~\ref{thm:H-sumrate}, we observe that a linear scaling can be achieved as $t \rightarrow \infty$ when the network
size $n$ grows faster than the constant term in the denominator of the packet throughput.
For a finite network size, however, this term cannot be neglected since it also grows with $t$.
Namely, adding more stages does not necessarily improve the achievable sum rate.
Thus, for given $n$, we can find the optimal number of hierarchical stages to maximize the sum rate.
In Theorem~\ref{thm:H-sumrate}, Methods 4, 2, 3, and 1 achieve the higher sum rate
in that order. Here, specifically, we considered Method 2 and find the optimal number of stages in closed form.
Since Method 1 is inferior to Method 2, we omit the corresponding result for the sake of brevity.
Unfortunately,  it is not possible to find a closed-form expressions for Methods 3 and 4. However, it is very easy to find the optimal number of stages
by a simple one-dimensional search for all the methods, and Fig.~\ref{opt_t} summarizes such results.
For Method 2,  we relax the integer constraint on $t$ and find the optimal $t$ as solution
of $\frac{d R_{{\rm sum}}^{(t)} (n,\alpha)}{ d t} = 0$.
This yields the equation
\[(t+1)^2\ln{\sqrt{3}} + (t+1) - \ln(n/L) = 0,\]
which provides the solution
\begin{equation}
t_{{\rm opt}} = -1+\frac{-1+\sqrt{1+2\ln(n/L)\ln{3}}}{\ln{3}}.\label{eq:topt}
\end{equation}
We notice that, even for $n$ as large as $10^{7}$, the optimal number of hierarchical stages
is not larger than 4 (see Fig.~\ref{opt_t}). Figs.~\ref{sum_routing} and~\ref{sum_routing_1} plot the achievable sum rates of the optimized hierarchical cooperation schemes
with an optimal number of stages. We observe that the proposed Method 4 provides a significant gain over the conventional schemes in \cite{Ozgur} and \cite{Ozgur3}, having a larger gain as $n$ increases. Remarkably, the optimized hierarchical cooperation schemes exhibit a sum rate that grows essentially linearly
with the number of users, for the considered range of networks of practical size.

\begin{remark} In \cite[Theorem 3.1]{Ghaderi}, the following refined achievable rate scaling law is determined as
\begin{align}
R_{{\rm sum}}^{(t)}(n,\alpha) = R\frac{n^{\frac{t}{t+1}}}{(t+1)(2\sqrt{1+Q/R})^t}\label{eq:similar}
\end{align} where $R$ and $Q$ are some unspecified constants, and the reuse factor $L$ is assumed to be $2$. Based on (\ref{eq:similar}), it was found in \cite{Ghaderi} that the optimal number of hierarchical stage is $t_{\rm opt} = O(\sqrt{\log{n}})$. In view of this previous work, one would be tempted to conclude that, for a sequence of networks of increasing size $n$, and optimal number of hierarchical cooperation layers
$t_{\rm opt}$ calculated as a function of $n$, the linear scaling of the sum rate is not achievable. In contrast,  we explicitly derive expressions for
the achievable sum rate by carefully optimizing the parameters of the hierarchical cooperation schemes by
using the TIN criterion as the guiding principle in order to obtain such system optimization in closed form.  As a consequence,
while on the basis of (\ref{eq:similar}) it is still impossible to make a quantitative comparison of hierarchical cooperation scheme versus multihop routing in terms of the actual achievable rate,
the results of Theorem~\ref{thm:H-sumrate} enable such a quantitative comparison, for networks of finite size.
We remark that optimizing the hierarchical cooperation schemes is not straightforward and, as a matter of fact, it was not done before.
In addition, we also observe that such optimization is necessary in order to make an accurate comparison with multihop routing. In fact, the standard hierarchical cooperation scheme of \cite{Ozgur} yields unsatisfactory performance, as shown in Figs.~\ref{sum_routing} and~\ref{sum_routing_1}.
Finally, based on our quantitative analysis (see Figs.~\ref{sum_routing} and~\ref{sum_routing_1}), we can conclude that hierarchical cooperation scheme can achieve a near-linear scaling of the sum rate in the range of networks of practical size.
\hfill $\lozenge$
\end{remark}

\begin{figure}
\centerline{\includegraphics[width=14cm]{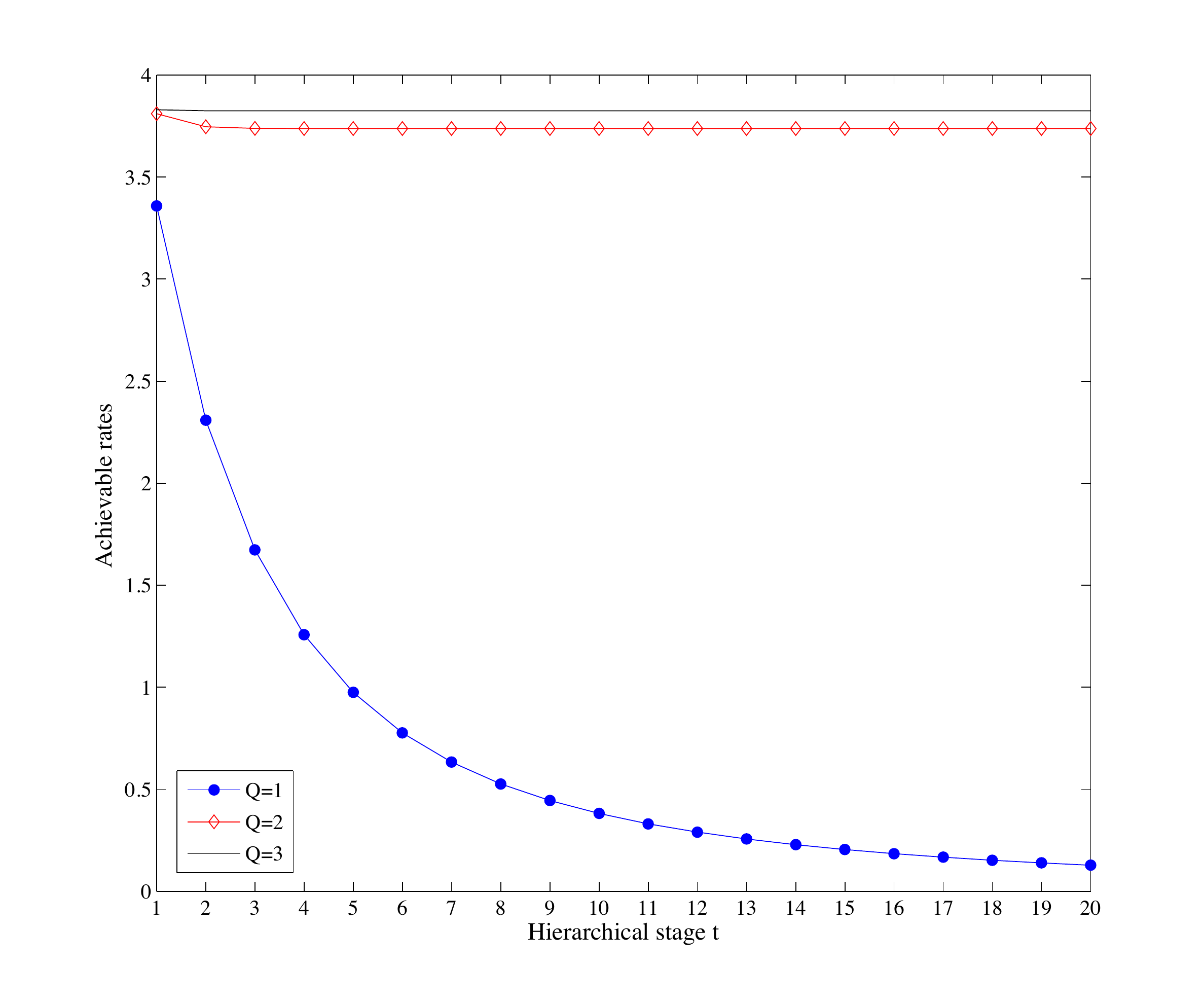}}
\caption{Achievable coding rate  as a function of hierarchical stage when pathloss exponent $\alpha=3$.}
\label{MRate}
\end{figure}

\subsection{Achievable coding rate}\label{subsec:M-RATE}

\begin{figure}
\centerline{\includegraphics[width=12cm]{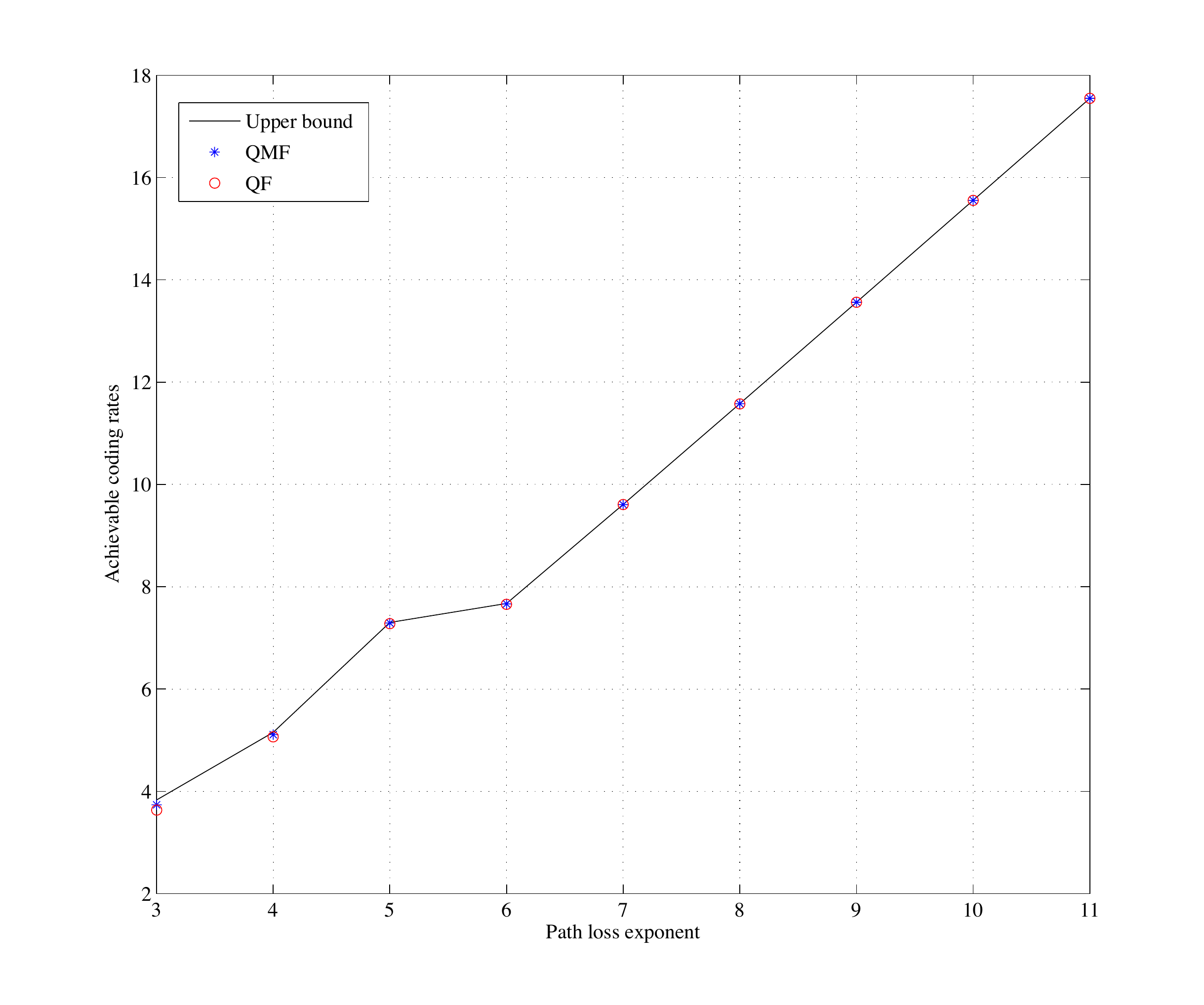}}
\caption{Achievable coding rate $R_{c}(\alpha)$ as function of pathloss exponent $\alpha$.}
\label{MRates}
\end{figure}

From Section~\ref{sec:CTS}, we have the rate-constraint of \[R_{{\rm c}}(\alpha) \leq R^{(1)} \eqdef \log\left(1+\frac{\SNR}{1+P_{I}}\right),\] for
reliable local communication at the bottom stage (i.e., stage 1).
Concatenating phases 2 and 3 of stage 1, we can produce an equivalent distributed MIMO channel
with backhaul capacity of $QR^{(1)}$ (see Section~\ref{subsec:MIMO}). Then, the coding rate should satisfy the
constraint
\[R_{{\rm c}}(\alpha) \leq R_{{\rm QMF}}(QR^{(1)},N_{0}=P_{I}+1,\SNR)  \eqdef R^{(2)}.\]
Lemma~\ref{lem:bound} below yields that $R^{(2)} \leq R^{(1)}$ for any positive integer $Q\geq 1$.
Since $R^{(2)}$ is the local communication rate of stage 2, we can produce a {\em degraded} distributed MIMO channel with backhaul
capacity $QR^{(2)} \leq QR^{(1)}$, resulting in the rate-constraint
\[R_{{\rm c}}(\alpha) \leq R_{{\rm QMF}}(QR^{(2)},N_{0}=P_{I}+1,\SNR)  \eqdef R^{(3)}.\]
Due to the smaller backhaul capacity,  we have that $R^{(3)} \leq R^{(2)}$.
Repeating the above procedure, we obtain that
\[R^{(t+1)} \eqdef  R_{{\rm QMF}}(QR^{(t)},P_{I}+1,\SNR) \leq R^{(t)},\]
implying that $\{R^{(t)}\}$ is monotonically non-increasing.
Hence, there exists a limit \[\lim_{t \rightarrow \infty } R^{(t)} = R^{\star}(\alpha,Q),\]
where such limit  depends on $\alpha$ and $Q$.
All rate-constraints are satisfied by choosing $R_{{\rm c}}(\alpha,Q) = R^{\star}(\alpha,Q)$.
One might have a concern that this choice is not a good one for small $t$.  However, Fig.~\ref{MRate} shows that $R^{(t)}$ quickly converges
to its positive limit for $Q\geq 2$.  Also, we observe that $Q=2$ is the best choice since it almost achieves the upper bound $R^{(1)}$,
while it requires only two time slots in order to deliver the quantization bits to the receivers in phase 3 of each stage.
Therefore, we choose $Q=2$ and $R_{{\rm c}}(\alpha) = R^{\star}(\alpha,2)$ in the following, for any $t$. The corresponding coding rates are plotted in Fig.~\ref{MRates} as a function of $\alpha$.

\begin{remark}
From Fig.~\ref{MRates}, we observe that QF almost achieves the upper bound for our model.
Thus, QF is better choice in practice, significantly reducing the decoding complexity at destination, since QMF requires
joint typical decoding based on the collected bin indices while QF just performs ``classical'' MIMO decoding based on the quantized
received signals (see comment in Section \ref{subsec:MIMO}).
\hfill $\lozenge$
\end{remark}

\begin{lemma}\label{lem:bound} For any $Q \geq 1$, the achievable rate of MIMO  transmission is upper-bounded by the local communication rate of bottom stage (i.e., stage 1):
\[R_{{\rm QMF}}(Q R^{(1)}, 1+P_{I}, \SNR) \leq R^{(1)}.\]
\end{lemma}
\begin{IEEEproof} Taking $Q \rightarrow \infty$, we have a naive upper bound:
\begin{equation*}
R_{{\rm QMF}}(QR^{(0)},P_{I}+1,\SNR) \leq R_{{\rm QMF}}(\infty, P_{I}+1,\SNR).
\end{equation*} Letting $\SNR_{{\rm eff}} =\frac{\SNR}{1+P_{I}}$, the proof follows from the limit upper bound:
\begin{eqnarray*}
R_{{\rm QMF}}(\infty,P_{I}+1,\SNR) &=& \lim_{M \rightarrow \infty} \frac{1}{M} \log\det\left(\Id+\frac{\SNR_{{\rm eff}}}{M}\Hm\Hm^{\herm}\right) \\
&\stackrel{(a)}{\leq}&\lim_{M \rightarrow \infty} \frac{1}{M}\log\left(\frac{\trace{\left(\Id+\frac{\SNR_{{\rm eff}}}{M}\Hm\Hm^{\herm}\right)}}{M}\right)^{M}\\
&\stackrel{(b)}{=}& \log(1+\SNR_{{\rm eff}})=R^{(1)},
\end{eqnarray*} where (a) is from the inequality of arithmetic and geometric means,  i.e., $\log\det(\Am) \leq M\log\left(\frac{\trace{(\Am)}}{M}\right)$ and (b) is due to the fact that $\lim_{M \rightarrow \infty} \frac{1}{M^2}\trace{(\Hm\Hm^{\herm})} = 1$ by Law of Large Numbers.
\end{IEEEproof}

\subsection{Achievable sum rate}\label{subsec:ST}

Focusing on the packet throughput, we review the previous works in \cite{Ozgur,Ozgur3} and propose an improvement by efficiently using the TDMA scheme during the local communication phases. For convenience, we let phase $(t,i)$ denote the phase $i$ of stage $t$.

\subsubsection{Method 1 (Conventional approach in \cite{Ozgur})}\label{subsubsection:approach1}

\begin{figure}
\centerline{\includegraphics[width=14cm]{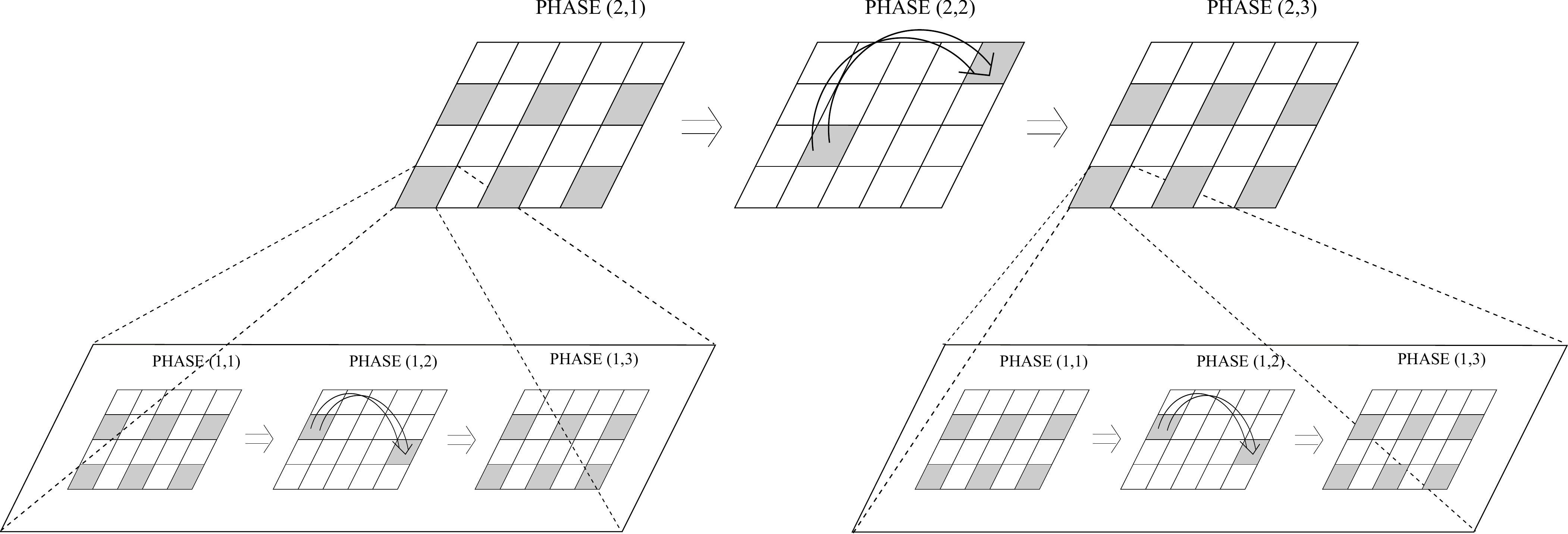}}
\caption{The silent features of the conventional schemes in \cite{Ozgur} when $2$-stage hierarchical cooperation architecture is applied.}
\label{eTDMA}
\end{figure}

The operation of stage 1 is equivalent to the cooperative transmissions discussed in Section~\ref{sec:CTS} and thus, from Section~\ref{subsec:ST1},
the packet throughput is given by
\begin{equation}
\mbox{T}^{(1)}(n,\alpha) = \frac{\sqrt{n}}{2L\sqrt{1+Q}}.\label{eq:TL1}
\end{equation}
Also, stage 2 employs stage 1 as its local communication both for information dissemination (phase (2,1)) and for cooperative reception (phase (2,3)) as shown in Fig.~\ref{eTDMA}.
Then, the required number of time slots is $(LM_{1})^2 / \mbox{T}^{(1)}(M_{1},\alpha)$ for phase (2,1), $n$ for phase (2,2), and
$Q(LM_{1})^2/\mbox{T}^{(1)}(M_{1},\alpha)$ for phase (2,3). The resulting packet throughput is given by
\[\mbox{T}^{(2)}(n,\alpha) = \frac{nM_{1}}{(1+Q)(LM_{1})^2/\mbox{T}^{(1)}(M_{1},\alpha) + n}.\] The optimal cluster size $M_{1}$ is obtained by solving $\frac{d \mbox{T}^{(2)}(n,\alpha)}{d M_{1}}=0$, which yields the \[M_{1} =\frac{n^{\frac{2}{3}}}{L^2(1+Q)}.\]
Then, we obtain the \[\mbox{T}^{(2)}(n,\alpha) =\frac{n^{\frac{2}{3}}}{3(L\sqrt{1+Q})^2}.\] Generalizing to $t$ stages, we obtain
the achievable sum rate of the method in \cite{Ozgur} as
\[R_{{\rm sum}}^{(t)} (n,\alpha) = R_{{\rm c}}(\alpha)\frac{n^{\frac{t}{t+1}}}{(1+t)(L\sqrt{1+Q})^{t}}.\]

\subsubsection{Method 2 (Improved TDMA scheduling)}\label{subsubsection:approach2}
\begin{figure}
\centerline{\includegraphics[width=14cm]{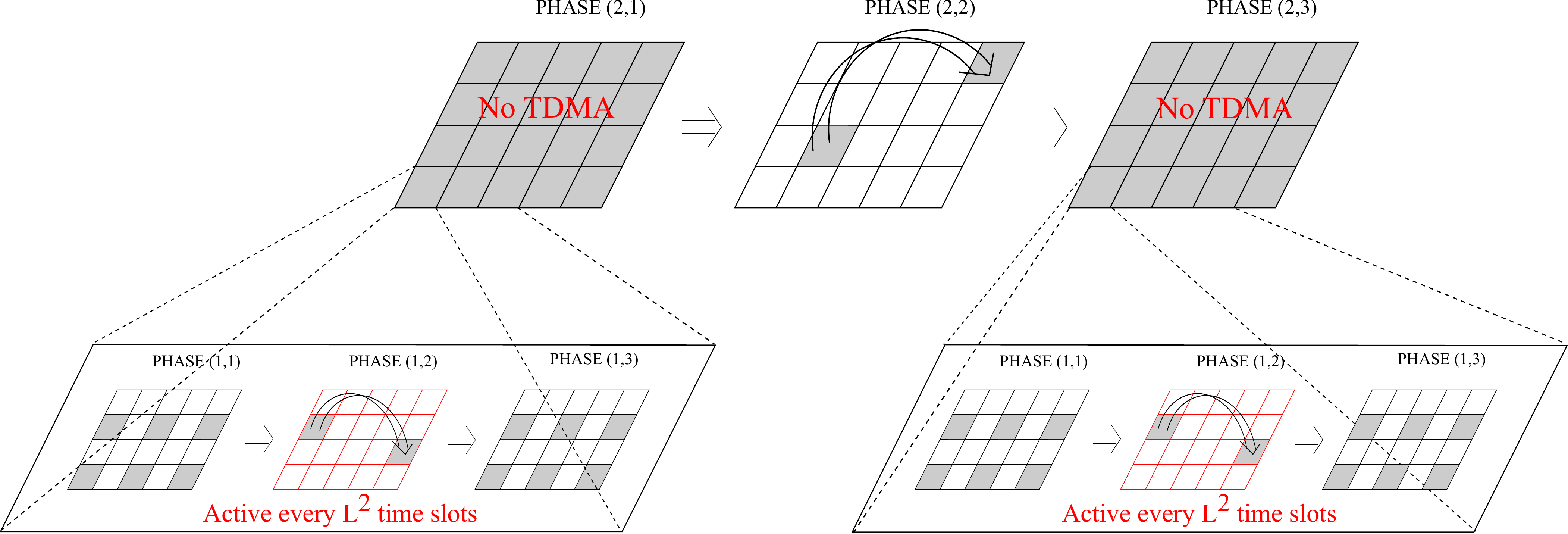}}
\caption{The silent features of the improved schemes when $2$-stage hierarchical cooperation architecture is applied.}
\label{eTDMA1}
\end{figure}

We will improve the penalty term associated with TDMA scheme from $L^{t}$ to $L^{\frac{2t}{t+1}}$.
This provides a non-trivial gain especially when $t$ is large, since the former exponentially increases with $t$
while the latter is upper bounded by $L^2$.
We explain our approach based on a 2-stage hierarchical cooperation architecture (see Fig.~\ref{eTDMA1}) and then
extend the result to general $t$. First, we remark that TDMA scheme is used so that the received power
of interference is less than a certain level for all transmissions. It can be noticed that this requirement is satisfied
for phases (1,1) and (1,3) without using the TDMA scheme of phase (2,1) since local communications have already
included the TDMA operation. However, the TDMA scheme of phase (2,1) is required for the long-range MIMO communication of phase (1,2).
Based on this observation, we can improve the basic scheme (Method 1) according to the scheme illustrated in Fig.~\ref{eTDMA1}:
All clusters in phase (2,1) (or phase (2,3)) are always active (spatial reuse 1);
In phase (1,2), each cluster has a turn to perform the MIMO transmissions every $L^2$ time slots,
which is equivalent to apply the TDMA scheme of phase (2,1) (or phase (2,3)). In short, this approach applies the TDMA scheme only once to
every phase. The resulting packet throughput can be computed as follows. Let $F_{1}(M)$ denote the number of time slots necessary to complete the local communication in each cluster in isolation, normalized by the message length in bits. Consider the local communication of phase (2,1) with $M$ nodes. Since TDMA is only applied to phase (1,2), it requires $F_{1}(M_{1}) + L^2 M + QF_{1}(M_{1})$ time slots to transmit $M M_{1}$ information messages, where $F_{1}(M) = L^2 M^2$ at the ground layer achieved by TDMA inside each cluster as shown in Fig.~\ref{eTDMA1}. Since there are $M^2$ transmissions in the local communication, we have
\begin{equation}
F_{2}(M) = \frac{M^2}{M M_{1}}(L^2 M + (1+Q)F_{1}(M_{1}))=\frac{M}{M_{1}}(L^2 M + (1+Q)F_{1}(M_{1})).
\end{equation} Then, the resulting packet throughput is given by
\begin{equation}
\mbox{T}^{(2)}(n,\alpha) = \frac{nM}{(1+Q)F_{2}(M)+n}
\end{equation} since there is no TDMA in the phases (2,1) and (2,3). Applying the above idea recursively, we obtain
\begin{equation}
F_{t}(n) = \frac{n}{M}(L^2 n + (1+Q)F_{t-1}(M)).
\end{equation} From Lemma~\ref{lem:NT}, we have:
\begin{equation}
F_{t}(n) = t L^2 (1+Q)^{\frac{t-1}{2}} n^{\frac{t+1}{t}}.\label{eq:Method2}
\end{equation} Using (\ref{eq:Method2}), we can compute the achievable sum rate of Method 2 as
\begin{align*}
R_{{\rm sum}} (n,\alpha) &= R_{c}(\alpha)\frac{nM_{t}}{(1+Q)F_{t}(M_{t}) + n}\nonumber\\
&\stackrel{(a)}{=}R_{c}(\alpha)\frac{nM_{t}}{tL^2(1+Q)^{\frac{t+1}{2}}M_{t}^{\frac{t+1}{t}} + n}\\
&\stackrel{(b)}{=}R_{c}(\alpha)\frac{n^{\frac{t}{t+1}}}{(1+t)L^{\frac{2t}{t+1}}(1+Q)^{\frac{t}{2}}},
\end{align*} where (a) is from (\ref{eq:Method2}) and (b) follows by choosing the optimal cluster size
\begin{equation}
M_{t} =\left( \frac{n}{L^2(1+Q)^{\frac{t+1}{2}}}\right)^{\frac{t}{t+1}}.\label{eq:optc1}
\end{equation}

%

\begin{lemma}\label{lem:NT} The number of time slots necessary to complete the local communication in each cluster in isolation, normalized by the message length in bits, is given by
\begin{equation}
F_{t}(n) = t L^2 (1+Q)^{\frac{t-1}{2}} n^{\frac{t+1}{t}}.
\end{equation}
\end{lemma}
\begin{IEEEproof} The result is provided by induction. The result holds for $t=1$. Assuming that it hods for $t-1$, we show that it also hods for $t$. We have:
\begin{align}
F_{t}(n) &= \frac{n}{M}(L^2 n (1+Q)F_{t-1}(M))\\
&=\frac{L^2n^2}{M}+(1+Q)^{\frac{t}{2}}(t-1)L^2 n M^{\frac{1}{t-1}}.\label{eq:lemma2-obj}
\end{align} We can minimize it by choosing the cluster size $M$ as solution of
\begin{equation}
\frac{d F_{t}(n)}{d M} = 0,
\end{equation}which yields the optimal cluster size as
\begin{equation}
M = \frac{n^{\frac{t-1}{t}}}{(1+Q)^{\frac{t-1}{2}}}.\label{eq:lemma2-optM}
\end{equation} By plugging (\ref{eq:lemma2-optM}) into (\ref{eq:lemma2-obj}), we can get:
\begin{equation}
F_{t}(n) = t L^2 (1+Q)^{\frac{t-1}{2}} n^{\frac{t+1}{t}}.
\end{equation}This completes the proof.
\end{IEEEproof}

\subsubsection{Method 3 (Network multiple access approach in \cite{Ozgur3})}\label{subsubsection:approach3} In the local communication (both phase 1 and phase 3) of the cooperative transmission scheme of Section~\ref{sec:CTS}, each node in the cluster wants to send independent messages to all other nodes in the same cluster. This communication problem is referred to as {\em network multiple access problem} in \cite{Ozgur3}. In Method 1 and Method 2, this problem is handled by decomposing it into a number of unicast network problems. In \cite{Ozgur3}, an improved communication method was presented as follows. Consider a network multiple access problem with $n$ nodes, where node has an $(n-1)$ independent information messages, one for each other node (all messages have the same size in bits). The strategy consists of dividing the network into cluster of size $M$. For each node $d$, and each cluster $\Sc$, all nodes in cluster $\Sc$ transmit their message destined to node $d$ at the same time. The $M$ nodes in cluster $\Dc$ (where $d$ belongs to), receive the superposition of these $M$ signals, quantize their received signals, and store them. Since we have $n$ destination nodes $d$, and each MIMO multiple access transmission carries the message of $M$ source nodes in $\Sc$, the number of such transmissions is $n^2/M$ (for each destination, we have one transmission per cluster). Now, with spatial reuse factor $L$, each cooperative reception cluster delivers its quantized received signal to each of its nodes. Notice that this local cooperative reception phase consists again of a network multiple access problem, on the subnetwork of size $M$ formed by a single cluster. Letting $F_{1}(M)$ denote the number of time slots necessary to complete the network multiaccess transmission in each cluster in isolation, normalized by the message length in bits, the time necessary to complete the cooperative transmission phase is $\frac{n}{M}L^2QF_{1}(M)$ (to disseminate information from each cluster $\Sc$, we have $L^2QF_{1}(M)$ transmissions) where, as before, $Q$ denotes the number of quantization bits. Hence, the time to complete the network multiaccess transmission for the whole network of $n$ nodes is given by
\begin{equation}
F_{2}(n) = \frac{n}{M}(n+L^2QF_{1}(M)).
\end{equation} Applying this idea recursively and letting $F_{1}(M) = M^2$ at the ground layer, achieved by TDMA inside each cluster, we obtain
\begin{equation}
F_{t}(n) = \frac{n}{M_{t}}(n+L^2 QF_{t-1}(M_{t})).\label{eq:recursive1}
\end{equation}  Using the recursive equation in (\ref{eq:recursive1}) and by finding an optimal cluster size $M_{t}$ (similar to what
done in Lemma~\ref{lem:NT}), we obtain
\begin{equation}
F_{t}(n) = t L^{t-1}Q^{\frac{t-1}{2}}n^{\frac{t+1}{t}}.\label{eq:NMAC}
\end{equation} Using (\ref{eq:NMAC}), we can compute the packet throughput as
\begin{equation}
T^{(t)}(n,\alpha) = \frac{nM}{L^2F_{t}(M)+n+L^2QF_{t}(M)}.\label{eq:NMAC2}
\end{equation} The optimal cluster size $M$ is obtained by solving $\frac{d T^{(t)}(n,\alpha)}{d M}=0$, which yields
\begin{equation}
M =L^{t}(1+Q)^{\frac{t}{t+1}}Q^{\frac{t(t-1)}{2(t+1)}} n^{\frac{t}{t+1}}.\label{eq:NMAC3}
\end{equation}Plugging (\ref{eq:NMAC3}) into (\ref{eq:NMAC2}), we have:
\begin{equation}
T^{(t)}(n,\alpha) = \frac{n^{\frac{t}{t+1}}}{(1+t)L^{t}\left((1+Q)Q^{\frac{t-1}{2}}\right)^{\frac{t}{t+1}}}.
\end{equation} In this method, the symmetric coding rate $R_{c}(\alpha)$ is determined by the rate constraints of degraded distributed MIMO channels as shown in Section~\ref{subsec:M-RATE}, since the concatenation of MIMO multiaccess transmission and cooperative reception is also considered as a distributed MIMO channel with a finite backhaul capacity and the backhaul capacity decreases as $t$ grows. Thus, Method 3 has the same symmetric coding rate of Method 1 and Method 2, and the achievable sum rate is given by
\begin{equation}
R_{{\rm sum}}^{(t)}(n,\alpha) = R_{c}(\alpha)\frac{n^{\frac{t}{t+1}}}{(t+1)L^{t}\left((1+Q)Q^{\frac{t-1}{2}}\right)^{\frac{t}{t+1}}}.
\end{equation} Comparing with Method 1, we observe that Method 3 enhances the penalty term associated with local communication from $(\sqrt{1+Q})^{t}$ to $\left((1+Q)Q^{\frac{t-1}{2}}\right)^{\frac{t}{t+1}}$.

\subsubsection{Method 4 (Combined approach of Method 2 and Method 3)}\label{subsubsection:approach4} We can further improve the achievable sum rate of Method 3 by using the efficient TDMA scheme as in Method 2. Namely, each cluster has a turn to perform the long-range MIMO multiaccess transmission every $L^2$ time slots and TDMA scheme is not applied for the cooperative reception phase. As explained in Section~\ref{subsubsection:approach2}, this approach guarantees that the received power of interference is less than a certain level for all transmissions. From this TDMA scheme, we obtain the following recursive equation:
\begin{equation}
F_{t}(n) = \frac{n}{M_{t}}(L^2 n+QF_{t-1}(M_{t})),\label{eq:recursive2}
\end{equation}where $F_{1}(n) = L^2 n^2$. Using the recursive equation in (\ref{eq:recursive2}) and by finding the optimal cluster size $M_{t}$, we obtain
\begin{equation}
F_{t}(n) = t L^{2} Q^{\frac{t-1}{2}}n^{\frac{t+1}{t}}.
\end{equation} Then, the packet throughput is given by
\begin{equation}
T^{(t)}(n,\alpha) = \frac{nM}{(1+Q)F_{t}(M)+n},
\end{equation} where notice that as in Method 2, the reuse factor $L^2$ is not applied to the long-range MIMO multiaccess transmission since one cluster at a time is active (i.e., no interference). By optimizing the cluster size $M$, we obtain the achievable sum rate of Method 4 as
\begin{equation}
R_{{\rm sum}}^{(t)}(n,\alpha) = R_{c}(\alpha) \frac{n^{\frac{t}{t+1}}}{(1+t)L^{\frac{2t}{t+1}}\left((1+Q)Q^{\frac{t-1}{2}}\right)^{\frac{t}{t+1}}}.
\end{equation} As expected, this method enhances the both penalties associated with TDMA and with local communication.

\section{Comparison with Multihop Routing}\label{sec:comp}

\begin{figure}
\centerline{\includegraphics[width=6cm]{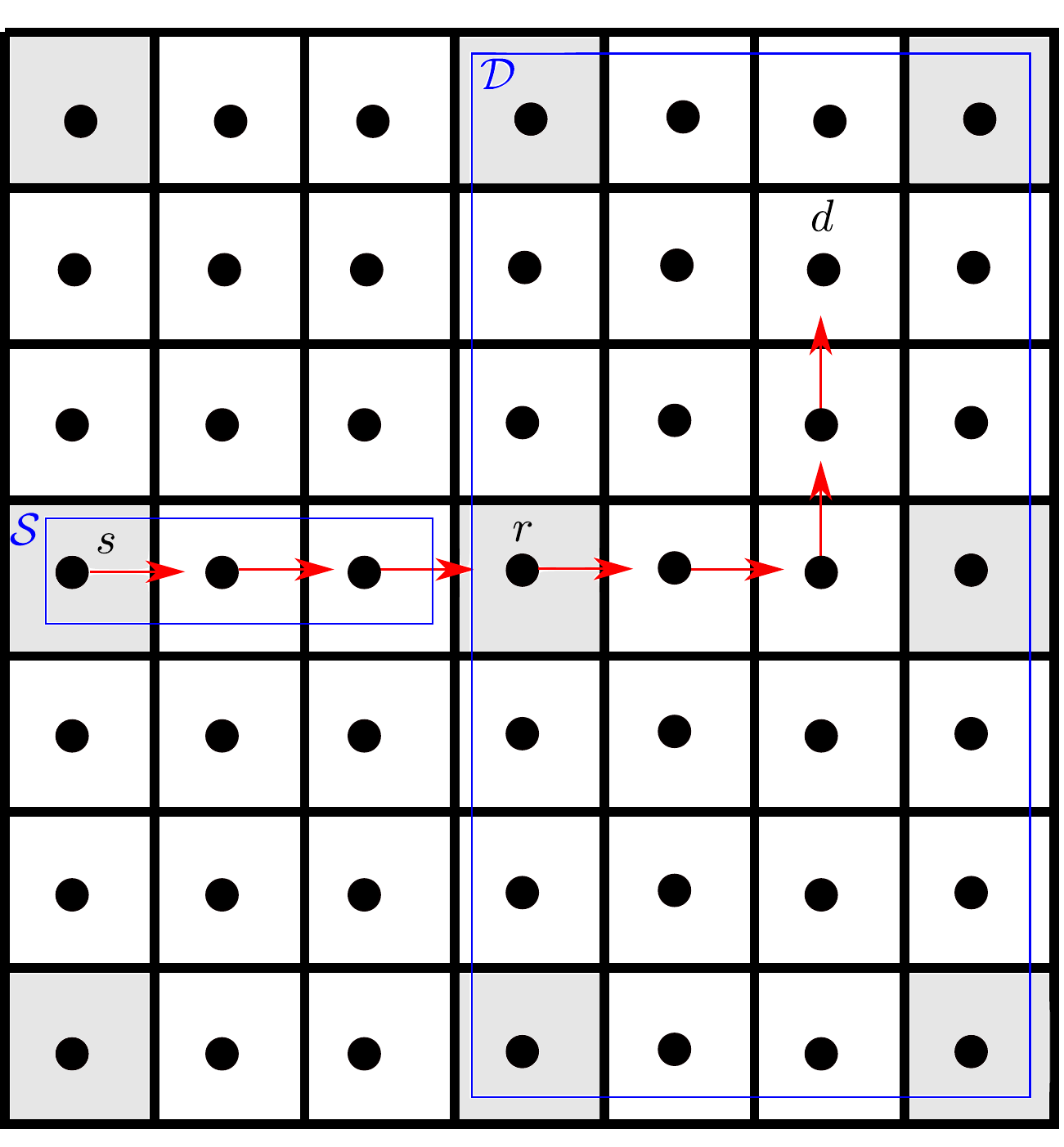}}
\caption{Multihop routing over the grid network. Red arrow-lines represent a routing path from the source $s$ to the destination $d$.}
\label{multi-hop}
\end{figure}

In this section, we compare the sum rate performance of our optimized hierarchical cooperation
scheme with that of conventional multihop routing (i.e., multihop decode and forward strategy).
In the multihop protocol, the packets of a source-destination pair are communicated by successive point-to-point transmissions
between relaying nodes. In \cite{Ozgur-book}, the performance of this scheme has been analyzed in terms of scaling law.
Also, it was shown that the cluster size $M=1$ maximizes the sum rate by minimizing the relaying burden.
Following this result, we assume $M=1$ and derive an achievable sum rate of multihop routing for the grid network considered in this paper.
Recall that the source-destination pairs are selected at random over the set of $n$-permutation $\pi$ that do not fix any element.
As in \cite{Ozgur-book}, we assume that the communication between each source-destination pair is relayed by the following
simple routing scheme: first proceeding horizontally and then vertically (see Fig.~\ref{multi-hop}). As done for local communication
in Section~\ref{sec:CTS}, the distance-dependent power control is applied and the interference is controlled by the reuse factor $L$,
chosen to enforce the optimality condition of TIN as $L(\SNR)=\left\lceil\sqrt{\SNR}^{1/\alpha}+1\right\rceil$.
The corresponding achievable coding rate for reliable hop transmission is given by
\begin{equation}
R_{{\rm c}}(\SNR) = \log\left(1+\frac{\SNR}{1+P_{I}}\right)
\end{equation} where $P_{I}$ is defined in (\ref{eq:IntPower}).

Next, we need to compute the packet throughput of multihop routing. The notation of Fig.~\ref{multi-hop} will be used in the following. Focusing on the center node $r$, the relaying traffic (i.e., the required number of time slots to deliver all messages) is generated either by the source nodes located in the same horizontal slab or the destination nodes located in the same vertical slab as $r$. The number of nodes contained in a slab is equal to $\sqrt{n}$. Then, the overall traffic generated by the source nodes located in the same horizontal slab as $r$ is upper bounded by $\sqrt{n}$. Also, the same computation can be applied to the overall traffic generated by the destination nodes located in the same vertical slab as $r$, which is bounded by $\sqrt{n}$.
Then, the overall average relaying traffic is equal to $2\sqrt{n}$. For other relay nodes, we can easily see that the traffic is not larger
than $2\sqrt{n}$. Including the impact of TDMA, the achievable sum rate of multihop routing is given by
\begin{equation}
R_{{\rm sum}}(n,\alpha) \geq \log\left(1+\frac{\SNR}{1+P_{I}}\right)\frac{\sqrt{n}}{2L^{2}(\SNR)}, \label{eq:MH-RATE}
\end{equation} where $L(\SNR) = \left\lceil\sqrt{\SNR}^{1/\alpha}+1\right\rceil$. Finally, we can find the optimal transmit power by differentiating and solving $\frac{d R_{{\rm sum}} }{d\SNR}=0$, yielding the optimal transmit power
\begin{equation}
\SNR = 2^{2(3+\alpha/(2\ln{2}))},
\end{equation}which is very close to the value found in Theorem \ref{thm:sum-rate}, for the single stage of the hierarchical cooperation architecture.

\begin{remark}({\em Average rate}) Since one can argue that it is conservative to use a lower bound for the performance comparison, we also compute the {\em average} relaying traffic. As before, we focus on the center node $r$. Let $\Sc=\{s_{1},\ldots,s_{|\Sc|}\}$ denote the set of source nodes located in the right side of $r$ (see Fig.~\ref{multi-hop}). Also, let $d_{i}$ denote the corresponding destination node for $i=1,\ldots,|\Sc|$. First, we compute the average relaying traffic generated by the source nodes in $\Sc$. This traffic is generated only when  $d_{i}$ is located in the right half-space, i.e., $d_{i} \in \Dc$ in Fig.~\ref{multi-hop}. Then, we
have
\begin{eqnarray}
\EE\left[\sum_{i=1}^{|\Sc|}\mbox{1}_{\{d_{i} \in \Dc\}}\right] &=& \sum_{i=1}^{|\Sc|}\PP(d_{i} \in \Dc)\\
&=& |\Sc|\frac{1}{2} = \frac{\sqrt{n}}{4},
\end{eqnarray}  where $\mbox{1}_{\{\Ec\}}$ denotes the indicator function  of an event $\Ec$ and $\PP(\cdot)$ is the probability measure
induced by the random source-destination assignment. In the above, we use the fact that $\PP(d_{i} \in \Dc) = \frac{1}{2}$ for
$i=1,\ldots,|\Sc|$ since for any given source node, its destination can be located in the right half-space with probability $\frac{1}{2}$.
With the exactly same argument, the average traffic generated by the source nodes located in the left side of $r$ is also equal to $\frac{\sqrt{n}}{4}$. Then, the overall average traffic generated by the source nodes located in the same horizontal slab as $r$  is equal to $\frac{\sqrt{n}}{2}$. Also, the same computation can be applied to the average traffic generated by the destination nodes located in the same vertical slab as $r$, which is equal to $\frac{\sqrt{n}}{2}$ (i.e., $\frac{\sqrt{n}}{4}$ generated from the destination nodes located in the upper side of $r$ and $\frac{\sqrt{n}}{4}$ generated from the destination nodes located in the lower side of $r$). Then, the overall average relaying traffic is equal to $\sqrt{n}$. We observe that average rate is two times higher than the lower bound as
\begin{equation}
R_{{\rm sum}}(n,\alpha) \geq \log\left(1+\frac{\SNR}{1+P_{I}}\right)\frac{\sqrt{n}}{L^{2}(\SNR)}, \label{eq:MH-RATE2}
\end{equation} where $L(\SNR) = \left\lceil\sqrt{\SNR}^{1/\alpha}+1\right\rceil$.
\hfill $\lozenge$
\end{remark}

\begin{remark} In this section, we compare the optimized hierarchical cooperation schemes and multihop routing for a network of size $n \leq 10^{5}$, for which
the optimal number of hierarchical stages $t_{{\rm opt}}$ is small ($t_{{\rm opt}} < 4$ for $n\leq 10^{5}$ from Fig.~\ref{opt_t}). When $t < 4$, it turns out that $Q=1$ can also guarantee a positive coding rate as shown in  Fig.~\ref{MRate}. Hence, the optimal $Q$ can be obtained by either $Q=1$ or $Q=2$, depending on the actual network size $n$ and and pathloss exponent $\alpha$. Taking this into account, we calculate the achievable rates of hierarchal cooperation schemes by maximizing over $t$ and $Q$, as
\begin{itemize}
\item Method 1:
\begin{equation}
R_{{\rm sum}}(n,\alpha) = \max_{t=1,\ldots,t_{{\rm max};Q=1,2}} R^{(t)} \frac{n^{\frac{t}{t+1}}}{(1+t)(L\sqrt{1+Q})^{t}} \label{eq:HC-RATE}
\end{equation}
\item Method 2:
\begin{equation}
R_{{\rm sum}}(n,\alpha) = \max_{t=1,\ldots,t_{{\rm max};Q=1,2}} R^{(t)} \frac{n^{\frac{t}{t+1}}}{(1+t)L^{\frac{2t}{t+1}}(1+Q)^{\frac{t}{2}}} \label{eq:HC-RATE2}
\end{equation}
\item Method 3:
\begin{equation}
R_{{\rm sum}}(n,\alpha) = \max_{t=1,\ldots,t_{{\rm max};Q=1,2}} R^{(t)} \frac{n^{\frac{t}{t+1}}}{(t+1)L^{t}\left((1+Q)Q^{\frac{t-1}{2}}\right)^{\frac{t}{t+1}}} \label{eq:HC-RATE3}
\end{equation}
\item Method 4:
\begin{equation}
R_{{\rm sum}}(n,\alpha) = \max_{t=1,\ldots,t_{{\rm max};Q=1,2}} R^{(t)} \frac{n^{\frac{t}{t+1}}}{(1+t)L^{\frac{2t}{t+1}}\left((1+Q)Q^{\frac{t-1}{2}}\right)^{\frac{t}{t+1}}}, \label{eq:HC-RATE4}
\end{equation}
\end{itemize} where the coding rate $R^{(t)}$ is defined in Section~\ref{subsec:M-RATE}, as a function of $\alpha$ and $Q$. Here, we employ a different coding rate depending on the number of hierarchical stages $t$, instead of using its limit as in Section~\ref{sec:H-CTS}, since for $Q=1$, $R^{(t)}$ is far from its limit for
large $t$. For the interesting network sizes (i.e., $n \leq 10^{5}$), the optimal number of hierarchical stages is small and the optimization in (\ref{eq:HC-RATE}) can be quickly solved by exhaustive search. 
\hfill $\lozenge$
\end{remark}

\begin{figure}
\centerline{\includegraphics[width=14cm]{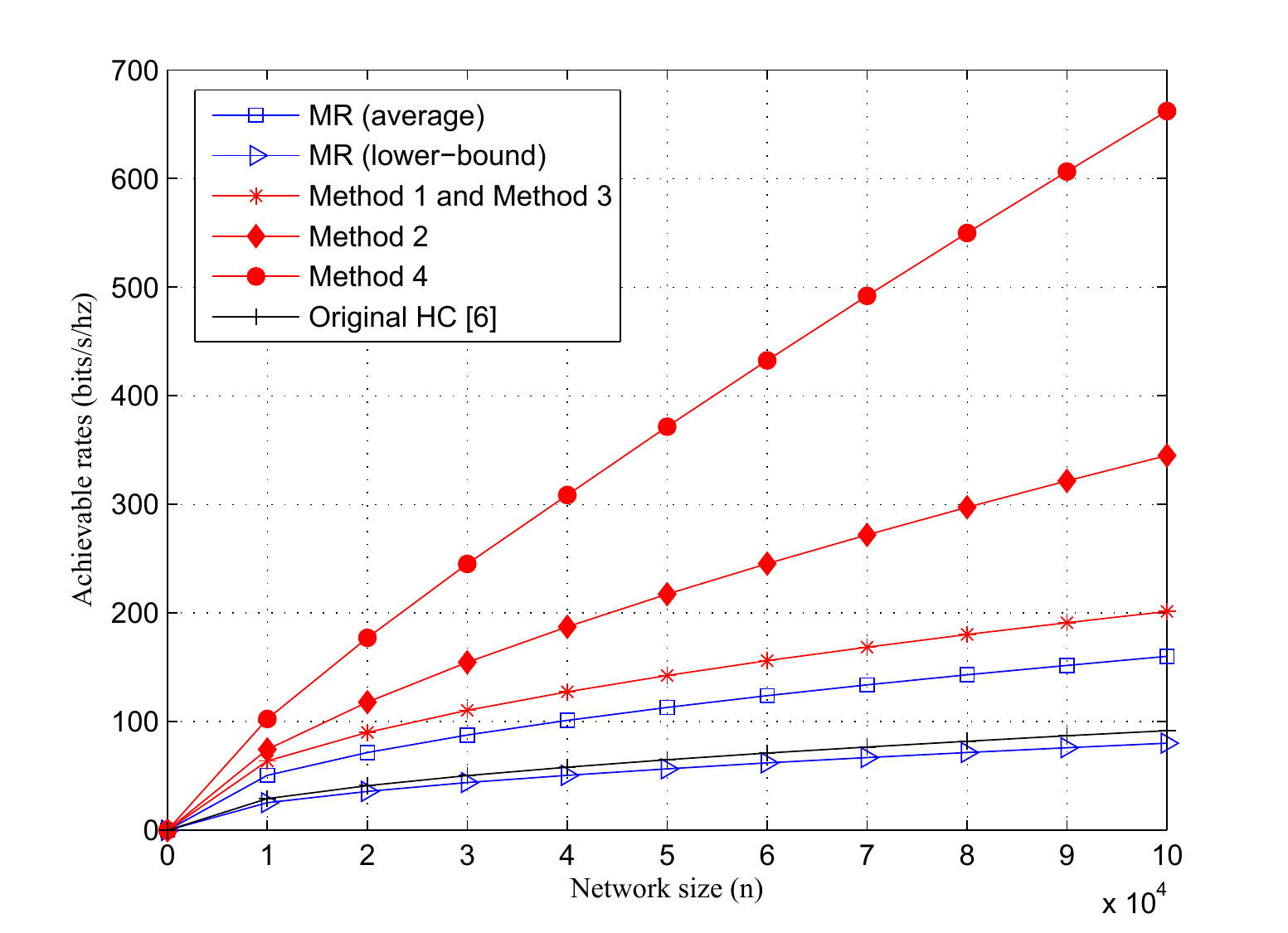}}
\caption{Performance comparison of hierarchical cooperation schemes and multihop routing when pathloss exponent $\alpha=7$. The original hierarchical cooperation scheme exactly follows the scheme in \cite{Ozgur} with $L=3$ and QF.}
\label{sum_routing}
\end{figure}

\begin{figure}
\centerline{\includegraphics[width=14cm]{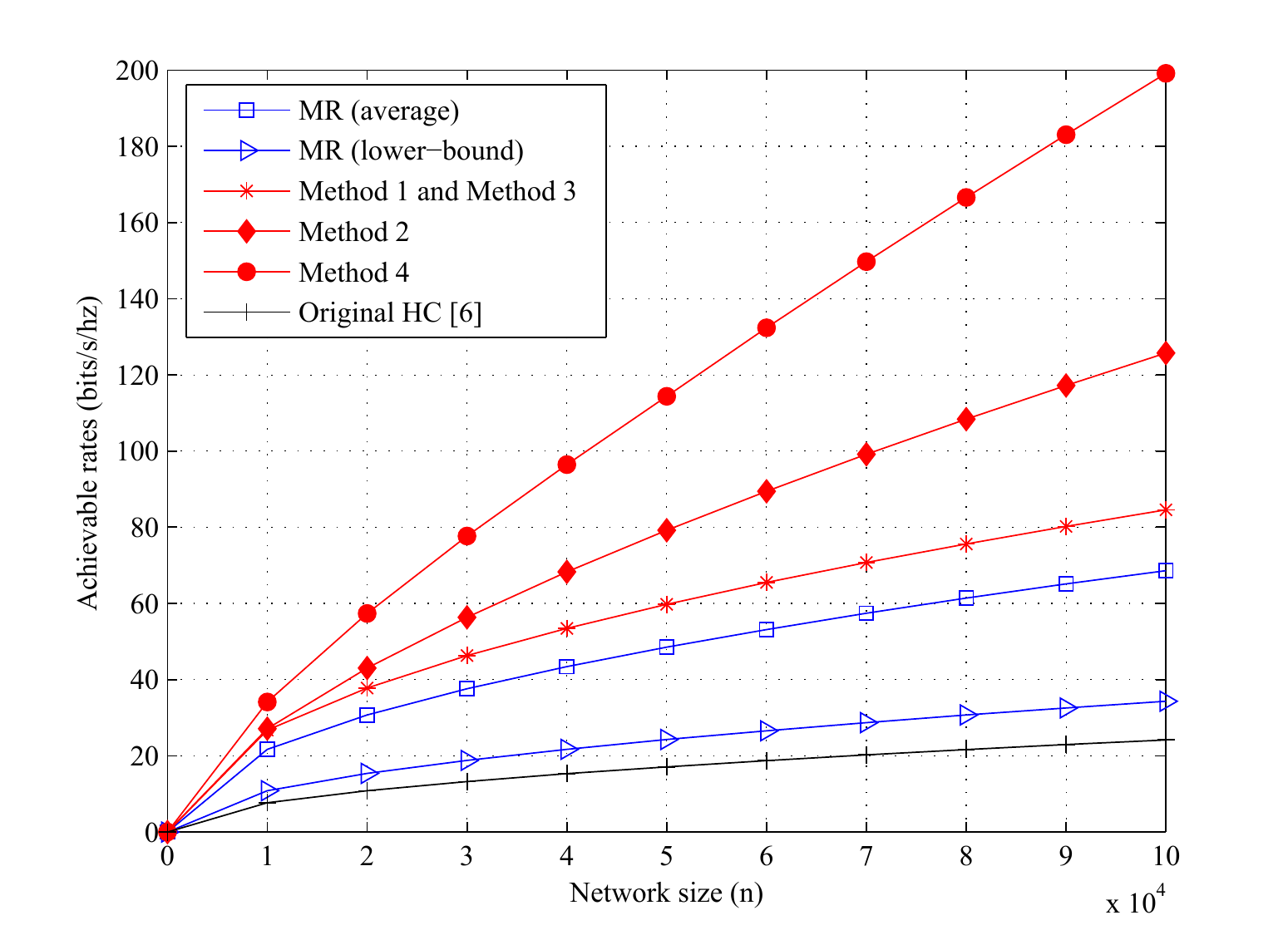}}
\caption{Performance comparison of hierarchical cooperation schemes and multihop routing when pathloss exponent $\alpha=4$. The original hierarchical cooperation scheme exactly follows the scheme in \cite{Ozgur} with $L=3$ and QF.}
\label{sum_routing_1}
\end{figure}

Fig.~\ref{sum_routing} and~\ref{sum_routing_1} plot the achievable sum rates of optimized hierarchical cooperation schemes in (\ref{eq:HC-RATE})-(\ref{eq:HC-RATE4}) and multihop routing in (\ref{eq:MH-RATE}) and (\ref{eq:MH-RATE2}). We observe that for $n\leq 10^5$, the optimal $Q$ of hierarchical cooperation schemes is equal to $1$, i.e., the phase 3 does not need to use time-expansion and transmit the quantization bits on multiple time slots. This is because for small network sizes (assuming that $n = 10^5$ can be considered as ``small''), the optimal number of hierarchical stages
is also small (i.e., not larger than 5) and hence the coding rate is large enough with $Q=1$ (as shown in Fig.~\ref{MRate}),
which minimizes the number of time slots. We also observe that when $Q=1$, QF cannot achieve the performance of QMF since the ``backhaul" capacity of phase 3 (without expansion) is not large enough. For example, when $\alpha=7$, Method 4 (with QMF) achieves the sum rate of $680$ bits/s/hz but Method 4 with QF achieves the sum rate of $396$ bits/s/hz. Also, we observe that the optimized hierarchical cooperation scheme
provides a higher sum rate than multihop routing, where the relative gain increases with the network size. Comparing the performances of Figs.~\ref{sum_routing} and~\ref{sum_routing_1}, we can see that the performance gains of the hierarchical cooperation schemes over multihop routing become larger as the pathloss exponent $\alpha$ grows. It is remarkable that, although the original hierarchical cooperation scheme without careful system parameters optimization does not provide significant gains over multihop routing as shown in Figs.~\ref{sum_routing} and~\ref{sum_routing_1},
the improved variants of the basic hierarchical cooperation scheme with our proposed system optimization yield significant gains in terms of achievable sum rate, and exhibit a sum rate that is a nearly linear function of the number of users $n$ in the range of networks of practical size $(n\leq 10^{5})$.

\begin{remark}
It is also interesting to assess whether the optimized hierarchical cooperation architecture is able to provide
attractive performances for future wireless networks.
Consider a network with area $\Ac=1\mbox{km}^2$ and number of users between $10^4$ and $10^5$ (this is
representative of a university or large industry campus).
Assume that the network operates in the mm-wave range, with carrier frequency $38$ GHz and bandwidth $200$ MHz (see \cite{Rapa}),
where $\alpha = 7$ can be expected. Fig.~\ref{sum_routing} shows that  the optimized hierarchical cooperation scheme can achieve a sum rate between 100 and $680$ bits/s/Hz,
such that the achievable  rate per source-destination pair is between 2 and 1.4 Mbps, respectively.
In the case of multihop routing, the sum rate varies from 25 to 80 bit/s/Hz and the corresponding achievable rate per source-destination pair
is between 500 and 160 Kbps, respectively.

It is also interesting to remark that if the same network operates in standard cellular bands (around 2GHz) with typical
pathloss $\alpha = 4$, and system bandwidth of 20 MHz, the rate per source-destination pair ranges from 68 to 40 Kbps
for the optimized hierarchical cooperation scheme, and from 20 to 3.5 Kbps for the multihop routing.

Given the significant gain of the optimized hierarchical cooperation scheme over multihop routing, and the fact that the network spectral efficiency is increases significantly with the pathloss exponent $\alpha$, we conclude that the combination of a carefully optimized hierarchical cooperation scheme and mm-wave communications may be a very attractive solution for future device-to-device dense wireless networks.
 \hfill $\lozenge$
\end{remark}

\section{Concluding Remarks}\label{sec:conclusion}

In this work we have taken a closer look at the actual achievable rate of hierarchical cooperation architectures for dense
device-to-device wireless networks. For the local communication phases, we focused on simple Gaussian codes, single-user decoding,
and treating interference as noise (TIN). We exploted the recent results on the approximate optimality of TIN \cite{Geng}, in order to
optimize the spatial reuse factor and the transmit power. It turns out that even though we assume an arbitrarily large  per-node power constraint
(as long as it is a fixed constant that does not scale with $n$), then the optimal transmit power is a constant value that depends only on
the pathloss exponent $\alpha$. For the global MIMO communication phase, we considered the optimization of the QMF approach
of \cite{Avestimehr}, observing that the combination of phase 2 and 3 of the MIMO cooperative scheme is formally analogous to the well-investigated
MIMO MAC channel with central processing and backhaul links of finite capacity. For such model, we have found new closed-form expressions for the
achievable rate in the case of large number of nodes and random i.i.d. channel coefficients, extending the formulas
provided for the symmetric Wyner model in \cite{Sanderovich}. Finally, we optimized further the achievable sum capacity by considering a variation of
the original hierarchical cooperation scheme in  \cite{Ozgur}, where we combine the TDMA phases of the hierarchical stages for better overall spectral efficiency.

The result of our optimization yields the performance of the hierarchical cooperation architecture in terms of actual achievable rates. We believe that
these rates cannot be easily beaten for this type of network model, as well as any network with random independent placements of the nodes
and random assignment of the source-destination pairs, for schemes that do not make ``non-physical assumptions'' on the communication
channel model (i.e., consider actual signal and noise power, and not artificial collision-based interference models such as the ``protocol model'')
and do not assume unreasonable knowledge of the network global state (i.e., this rules out interference alignment schemes based on
the knowledge of the network state with infinite precision). Furthermore, since the scheme considered here involves only
``Gaussian'' single-user coding, the analysis of this paper is suitable to be extended to more refined finite-length analysis
\cite{Poly}, where the tradeoff between coding length and block error probability can be also investigated, thus illuminating also
issues about the latency of such networks. This interesting aspect, however, is out of the scope of the present paper and it is
deferred to future work.

The main conclusions of our work are summarized as follows.
 Despite in terms of sum rate scaling, when we take the limit for large $n$ and optimize for each $n$ the number of hierarchical stages $t$,
the linear scaling of the sum rate versus the number of users $n$ seems not attainable (in agreement with previous findings in \cite{Ghaderi}),
our quantitative analysis shows that for a wide range of practically relevant network sizes and in realistic conditions of SNR and pathloss exponent
the improved and optimized hierarchical cooperation architecture developed in this paper achieves a nearly linear scaling of the sum rate. Remarkably,
this is so even if the optimal number of hierarchical stages is not not larger than 4, even for unreasonably large networks (up to $n = 10^7$).
Furthermore, we were able to compare the sum rate (not just the scaling law) achievable by hierarchical cooperation scheme with that achievable by conventional multihop routing.
To the best of our knowledge, this is the first time that such quantitative comparison is made.
Thanks to our quantitative analysis, we were able to shed light on the long-standing question of whether hierarchical cooperation scheme
can yield throughput advantages over multihop routing in terms of actual user rates, in practically relevant network scenarios.
We found that, in a scenario representative of a possible next-generation device-to-device network operating at mm-waves,
about 800\% (a factor of 8) rate gain of hierarchical cooperation scheme over multihop routing can be expected.

Another more subtle aspect that we wish to just mention here is that,
since for practical network sizes the optimal number of hierarchical stages $t$ is small, then
the optimized hierarchical cooperation scheme delivers the messages to their destination in a small number
of hops (including both local and global MIMO hops). Instead, multihop routing yields on average $\Theta(\sqrt{n})$ number of hops
to deliver messages. This may yield a significant difference in latency and in coding rate when considering the effect of finite
block length and non-zero block error rate. This interesting question is left for future investigation.

\section*{Acknowledgment}

This work was supported by NSF Grant CCF 1161801.

\appendices
\section{Achievable rate of Distributed MIMO channel with finite backhaul capacity }\label{sec:DAS-RATE}
In this section, we derive a closed-form expression of achievable rate of QMF for the distributed MIMO channel with finite backhaul capacity of rate $R_{0}$ (see Fig.~\ref{DAS}).
Let  $\Hm \in \CC^{M \times M}$  denote the channel matrix,  with $k$-th row $\hv_{k}$, for $k=1,\ldots,M$, and i.i.d. elements
with zero mean and unit variance. Also, we assume that the transmit power of each user is equal to $\frac{\SNR}{M}$ (i.e., the sum-power is fixed as $\SNR$) and we let $N_{0}$ denote the total received interference plus noise power.  We focus on the {\em symmetric} user rate.
The following notations will be frequently used in this section. Let $\xv$ and $\yv$ denote the $M$-dimensional transmit and receiver vectors,
respectively. Let $\Sc \subseteq [1:M]$ denote the row index set of $\Hm$.  For given $\Sc \subseteq [1:M]$, $\Hm_{\Sc}$ represents the
channel sub-matrix of the inputs $\xv$ to the outputs $\yv_{\Sc}$.

This model has been extensively studied in \cite{Sanderovich,Hong}. Interested readers should refer to \cite{Sanderovich,Hong} for some discussion and comparison of various relaying schemes. The cut-set upper bound of such channel is given by
\begin{eqnarray*}
R_{\rm{upper}}=\frac{1}{M} \min_{\Sc \subseteq [1:M]} |\Sc|R_{0} + \log\det\left(\Id+\frac{\SNR}{N_{0}}\frac{\Hm_{\Sc^{c}}\Hm_{\Sc^{c}}^{\herm}}{M}\right).
\end{eqnarray*} This result follows by a considering cut-set bound for two cuts: one obtained by separating the destination from the
receivers and the other obtained by separating the receivers from the transmitters.
The achievable rate of QMF (with given quantization distortion level $\sigma_{q,i}^2$) is derived in \cite{Sanderovich} as
\begin{eqnarray}
R_{\rm QMF} &=&\frac{1}{M} \min_{\Sc \subset [1:M]} \sum_{i \in \Sc} \left(R_{0}- \log\left(1+\frac{N_{0}}{\sigma_{q,i}^2}\right)\right)+ \log\det\left(\Id+\diag\left(\frac{\SNR}{N_{0}+\sigma_{q,i}^2}\right)\frac{\Hm_{\Sc^{c}}\Hm_{\Sc^{c}}^{\herm}}{M}\right).\nonumber \\ \label{eq:QMFrate}
\end{eqnarray}
Also, Quantize and Forward (QF), a simplified version of QMF that does not include binning after quantization, and directly forwards the quantization bits to the central receiver, achieves the rate of
\begin{eqnarray*}
R_{{\rm QF}}
&=&\frac{1}{M} \log\det\left(\Id+\diag\left(\frac{(2^{R_{0}}-1)\SNR}{2^{R_{0}}N_{0}+\SNR'\|\hv_{i}\|^2}\right)\frac{\Hm\Hm^{\herm}}{M}  \right)
\end{eqnarray*} which is obtained from (\ref{eq:QMFrate}) by setting the quantization level as
\begin{equation}
\sigma_{q,i}^2=\frac{N_{0}+\SNR'\|\hv_{i}\|^2}{2^{R_{0}}-1} \mbox{ for }i=1,\ldots,M.\label{eq:QFlevel}
\end{equation}

Next, we will derive the closed-form expressions of the above rates in order to prove Theorem~\ref{thm:DASrate}.
This will be obtained from asymptotic Random Matrix Theory results, and using the submodular structure of the rate expression.
We first provide some lemmas that will be used to prove the theorem.

\begin{definition}\label{def:submodular} Let $\Omega=[1:M]$ be a finite ground set. A set function $f:2^{\Omega} \rightarrow \RR$ is {\em submodular} if for every set $\Ac,\Bc \subseteq \Omega$ with $\Ac \subseteq \Bc$ and every $x \notin \Bc$, the following is satisfied:
\begin{equation}
f(\Ac \cup \{x\}) - f(\Ac) \geq f(\Bc \cup \{x\}) - f(\Bc).
\end{equation}
\hfill $\lozenge$
\end{definition}
Intuitively, submodular functions capture the concept of diminishing returns: as the set becomes larger the benefit of adding a new element to the set will decrease.
\begin{lemma}\label{lem:conc} Suppose that a set function $f(\Sc)$ only depends on the size of subset $|\Sc|$, i.e., for any $\Sc_{1},\Sc_{2} \subseteq [1:M]$
\begin{equation}
f(\Sc_{1}) = f(\Sc_{2})  \mbox{ if } |\Sc_{1}|=|\Sc_{2}|.
\end{equation} Define
\[g(\beta)\eqdef f(\Sc)\] where $0\leq \beta=\frac{|\Sc|}{M}\leq 1$. If $f(\Sc)$ is submodular, then $g(\beta)$ is {\em concave}
when $M \rightarrow \infty$.
\end{lemma}
\begin{IEEEproof} Since $f(\Sc)$ is submodular, the following inequality holds from Definition~\ref{def:submodular}:
\begin{equation}
f(\Ac \cup \{x\}) - f(\Ac) \geq f(\Bc \cup\{x\}) - f(\Bc) \label{eq:submodular}
\end{equation} for any $\Ac\subset \Bc \subset [1:M]$
and $x \notin \Bc$. From (\ref{eq:submodular}) and the assumption of $f(\Sc)$ only depending on the size of subset $|\Sc|$, the following inequality also holds for any $\beta$ and $\beta'$ with $\beta' > \beta$:
\begin{equation}
g(\beta + \Delta) - g(\beta) \geq g(\beta' + \Delta) - g(\beta')\label{eq:der}
\end{equation} where $\Delta = \frac{1}{M}$.  Letting $M \rightarrow \infty$, the (\ref{eq:der}) implies that  $\dot{g}(\beta) \geq \dot{g}(\beta')$ for any $\beta
< \beta'$, i.e., $\dot{g}(\beta)$ is monotonically decreasing function. Therefore, $g(\beta)$ is a concave function.
\end{IEEEproof} The following is the main result of this section, which completes the proof of Theorem~\ref{thm:DASrate}.

\begin{lemma} In the large system limit (i.e., $M \rightarrow \infty$), we have:
\begin{eqnarray*}
R_{\rm{upper}}&=& \min\left\{R_{0},\Cc\left(\frac{\SNR}{N_{0}}\right)\right\}\\
R_{\rm{QMF}} &=& \min\left\{R_{0}-\log\left(1+\frac{N_{0}}{\sigma_{q}^2}\right),\Cc\left(\frac{\SNR}{N_{0}+\sigma_{q}^2}\right)\right\}\\
R_{\rm{QF}} &=& \Cc\left(\frac{(2^{R_{0}}-1)\SNR}{2^{R_{0}}N_{0}+\SNR}\right)
\end{eqnarray*} where $\Cc(\cdot)$ is given by (\ref{Cx}).
\end{lemma}
\begin{IEEEproof} We only prove the closed-form expression of cut-set upper bound since the others are straightforwardly proved
along the same lines.  Letting $\beta =\frac{|\Sc^{c}|}{M}$, we have:
\begin{eqnarray*}
R_{\rm{upper}}\nonumber &=&\lim_{M \rightarrow \infty} \min_{\Sc \subset [1:M]} \frac{|\Sc|}{M}R_{0} + \frac{1}{M} \log\det\left(\Id+\frac{\SNR}{N_{0}}\frac{\Hm_{\Sc^{c}}\Hm_{\Sc^{c}}^{\herm}}{M}\right)\nonumber\\
&\stackrel{(a)}{\rightarrow}& \min_{\Sc \subseteq [1:M]} (1-\beta)R_{0} + \Cc\left(\frac{\SNR}{N_{0}},\beta\right) \mbox{ as $M\rightarrow \infty$}\nonumber\\
&=& \min_{0\leq \beta \leq 1} (1-\beta)R_{0} +\Cc\left(\frac{\SNR}{N_{0}},\beta\right)\label{eq:min}\\
&\stackrel{(b)}{=}& \min\left\{R_{0},\Cc\left(\frac{\SNR}{N_{0}}\right)\right\}.
\end{eqnarray*}
\begin{itemize}
\item The (a) is obtained from asymptotic Random Matrix Theory in \cite{Verdue}, given by
\begin{equation*}
\frac{1}{M}\log\det\left(\Id + \frac{\SNR}{M}\Hm_{\Sc^{c}}\Hm_{\Sc^{c}}^{\herm}\right) \rightarrow \Cc(\SNR,\beta) \mbox{ as } M \rightarrow \infty
\end{equation*} where
\begin{eqnarray*}
\Cc(\SNR,\beta)&=& \beta\log\left(1+\frac{\SNR}{\beta}-\frac{1}{4}\Fc\left(\frac{\SNR}{\beta},\beta\right)\right) +\log\left(1+\SNR-\frac{1}{4}\Fc\left(\frac{\SNR}{\beta},\beta\right)\right) \\ &&-\beta \frac{\log{e}}{4\SNR}\Fc\left(\frac{\SNR}{\beta},\beta\right)\\
\Fc(x,z) &=& \left(\sqrt{x(1+\sqrt{z})^2+1}-\sqrt{x(1-\sqrt{z})^2+1}\right)^2.
\end{eqnarray*}
\item The (b) is due to the fact that $f(\Sc^{c}) \eqdef  \log\det\left(\Id+\frac{\SNR}{N_{0}}\frac{\Hm_{\Sc^{c}}\Hm_{\Sc^{c}}^{\herm}}{M}\right)$ is a submodular \cite{JB85} and hence, from Lemma~\ref{lem:conc}, the $\Cc\left(\frac{\SNR}{N_{0}},\beta\right)$ is a concave function. Thus, the minimum of (\ref{eq:min}) is
attained at the boundary, i.e., either for $\beta=0$ or for $\beta=1$. Also, we use the simple notation as $\Cc(\SNR,1) = \Cc(\SNR)$.
\end{itemize} With the same arguments, we can prove the closed-form expressions of achievable rates of QMF and QF.
\end{IEEEproof}



\begin{thebibliography}{1}

\bibitem{Avestimehr} S. Avestimehr, S. Diggavi, and D. Tse, ``Wireless network information flow: A deterministic approach," {\em IEEE Transactions on Information Theory,} vol. 57, pp. 1872-1905, Apr. 2011.

\bibitem{Cadambe08} V. Cadambe and S. A. Jafar, ``Interference alignment and the degrees of freedom of the K user interference channel," {\em IEEE Transactions on Information Theory,} vol. 54, pp. 3425-3441, Aug. 2008.

\bibitem{Gou09} T. Gou and S. A. Jafar, ``Capacity of a class of symmetric SIMO Gaussian interference channels within O(1)," in {\em Proceedings of IEEE International Symposium on Information Theory (ISIT),} Seoul, Korea, Jun-Jul. 2009.

\bibitem{Jafar10} S. A. Jafar and S. Vishwanath, ``Generalized Degrees of Freedom of the Symmetric Gaussian K User Interference Channel," {\em IEEE Transactions on Information Theory,} vol. 56, pp. 3297-3303, Jul. 2010.



\bibitem{Gupta} P. Gupta and P. R. Kumar, ``The capacity of wireless networks," {\em IEEE Transactions on Information Theory,} vol. 46, pp. 388-404, Mar. 2000.


\bibitem{Ozgur} A. Ozgur, O. Leveque, and D. Tse, ``Hierarchical Cooperation Achieves Optimal Capacity Scaling in Ad Hoc Networks," {\em IEEE Transactions on Information
    Theory,} vol. 53, pp. 3549-3572, Oct. 2007.

\bibitem{Franceschetti} M. Franceschetti, M. D. Migliore, and P. Minero, ``The capacity of wireless networks: Information-theoretic and physical limits," {\em IEEE Transactions on Information Theory,} vol. 55, pp. 3413-3424, Aug. 2009.


\bibitem{Lee} S.-H. Lee and S.-Y. Chung, ``Capacity scaling of wireless ad hoc networks: Shannon meets Maxwell," {\em IEEE Transactions on Information Theory,} vol. 58, pp. 1702-1715, Mar. 2012.


\bibitem{Ozgur1} A. Ozgur, O. Leveque, and D. Tse, ``Spatial degrees of freedom of large distributed MIMO systems and wireless ad hoc networks," {\em IEEE Journal on Selected Areas in Communication,} vol. 31, pp. 202-2014, Feb. 2013.


\bibitem{Ghaderi} J. Ghaderi, L.-L. Xie, and X. Shen, ``Hierarchical Cooperation in Ad Hoc Networks: Optimal Clustering and Achievable Throughput," {\em IEEE Transactions on Information Theory,} vol. 55, pp. 3425-3436, Aug. 2009.

\bibitem{Daniels} R. Daniels, R. Heath, J. Murdock, and T. Rappaport, {\em 60 GHz Wireless Communication Systems} Prentice Hall Press, 2012.

\bibitem{rappa} T. Rappaport, F. Gutierrez, E. Ben-Dor, J. Murdock, Y. Qiao, and J. Tamir, ``Broadband Millimeter-Wave Propagation Measurements and Models Using Adaptive-Beam Antennas for Outdoor Urban Cellular Communications," {\em IEEE Transactions on Antennas and Propagation,} vol. 61, pp. 1850-1859, 2013.

\bibitem{rappa-mag}
T. Rappaport, S. Sun, R. Mayzus, Hang Zhao, Y. Azar, K. Wang, G. N. Wong, J. K. Schulz, M. Samimi, M. and F. Gutierrez,
``Millimeter Wave Mobile Communications for 5G Cellular: It Will Work!,''
{\em Access, IEEE}, Vol. 1, pp. 335 - 349, 2013.



\bibitem{Ozgur3} A. Ozgur and O. Leveque, ``Throughout-Delay Tradeoff for Hierarchical Cooperation in ad hoc wireless networks," {\em IEEE Transactions on Information Theory,} vol. 56, pp. 1369-1377, Mar. 2010.

\bibitem{Lim} S. Lim, Y. H. Kim, A. E. Gamal, and S. Chung, ``Noisy Network Coding," {\em IEEE Trans. Inf. Theory,} vol. 57, pp. 3132-3152.



\bibitem{Xie} L. L. Xie, ``On Information-Theoretic Scaling Laws for Wireless Networks," http://arxiv.org/pdf/0809.1205.pdf






\bibitem{Sanderovich} A. Sanderovich, O. Somekh, H. V. Poor, and S. Shamai (Shitz), ``Uplink Macro Diversity of Limited Backhaul Cellular Network," {\em IEEE Transactions on Information Theory,} vol. 55, pp. 3457-3478, Aug. 2009.


\bibitem{Yanikomeroglu} H. Yanikomeroglu, ``Cellular multihop communications: Infrastructure-based relay network architecture 4G wireless systems," in {\em proceedings of 22nd Bienniel Symposium on Communications,} Ontario, Canada, Jun. 2004.


\bibitem{Han} T. S. Han and K. Kobayashi, ``A New Achievable Rate Region for the Interference Channel," {\em IEEE Transactions on Information Theory,} vol. 27, pp. 49-60, Jan. 1981.

\bibitem{Etkin} R. H. Etkin, D. N. C. Tse, and H. Wang, ``Gaussian Interference Channel Capacity to Within One Bit," {\em IEEE Transactions on Information Theory,} vol. 54, pp. 5534-5562, Dec. 2008.

\bibitem{Cadambe} V. R. Cadambe and S. A. Jafar, ``Interference Alignment and Degrees of Freedom of the K-User Interference Channel," {\em IEEE Transactions on Information Theory,} vol. 54, pp. 3425-3411, Aug. 2008.









\bibitem{Ordentlich} O. Ordentlich, U. Erez, and B. Nazer, ``The Approximate Sum Capacity of the Symmetric Gaussian K-User Interference Channel," [Online] arXiv:1206.0197.



\bibitem{Geng} C. Geng, N. Naderializadeh, A. S. Avestimehr, and S. A. Jafar, ``On the Optimality of Treating Interference as Noise," [Online] http://arxiv.org/abs/1305.4610.





\bibitem{Hong} S.-N. Hong and G. Caire, ``Compute-and-Forward Strategies for Cooperative Distributed Antenna Systems," {\em IEEE Transactions on Information Theory,} vol. 59, pp. 5227-5243, Aug. 2013.


\bibitem{Nazer} B. Nazer and M. Gastpar, ``Compute-and-Forward: Harnessing Interference through Structured Codes,''\emph{IEEE Transactions on Information Theory}, vol. 57, pp. 6463-6486, Oct. 2011.

\bibitem{Chern} B. Chern and A. Ozgur, ``Achieving the capacity of the N-relay Gaussian diamond network within $\log{n}$ bits," {\em IEEE Transactions on Information Theory,} vol. 60, pp. 7708-7718, Dec. 2014.

\bibitem{Kolte} R. Kolte and A. Ozgur, ``Improved capacity approximations for Gaussian relay networks," in {\em proceedings of IEEE Information Theory Workshop (ITW),} pp. 1-5, Sep. 2013

\bibitem{Hong-full} S.-N. Hong and G. Caire, ``Full-Duplex Relaying with Half-Duplex relays," {\em submitted to IEEE Trans. Inf. Theory,} 2013.

\bibitem{Hong-ITW2015} S.-N. Hong, I. Maric, D. Hui, and G. Caire, ``Multihop Virtual Full-Duplex RElay Channels," in {\em proceedings of IEEE Information Theory Workshop (ITW),} Apr.-May 2015.

\bibitem{Ozgur-book} A. Ozgur, O. Leveque, and D. Tse, ``Operating Regimes of Large Wireless Networks," {\em Foundations and Trends in Networking,} 2011.

%
\bibitem{Rapa} T. Rappaport, F. Gutierrez, E. Ben-Dor, J. Murdock, Y. Qiao, and J. Tamir, ``Broadband millimeter wave propagation measurements and models using adaptive beam antennas for outdoor urban cellular communications," 2011.


\bibitem{Poly} Y. Polyanskiy, H. V. Poor and S. Verdue, ``Channel coding rate in the finite blocklength regime," {\em IEEE Transactions on Information Theory,} vol. 56, pp. 2307-2359, May 2010.


\bibitem{Verdue}  S. Verdu and S. Shamai, ``Spectral efficiency of CDMA with random spreading," {\em IEEE Transactions on Information Theory,} vol. 45, pp. 622-640, Mar. 1999.
%
%
%
%
\bibitem{JB85} C. R. Johnson and W. W. Barrett, ``Spanning-tree extensions of the Hadamard-Fischer inequalities," {\em Linear Algebra Applications}, pp. 177-193, 1985.

\end{thebibliography}
\end{document}